\documentclass[11pt,a4paper]{article}
\usepackage[utf8]{inputenc}
\usepackage[T1]{fontenc}
\usepackage[numbers, sort&compress]{natbib}
\usepackage[french]{babel}
\usepackage{amsmath,amssymb,amsthm}
\usepackage{geometry}
\usepackage{lmodern}
\usepackage{mathtools}
\usepackage{microtype}
\usepackage{graphicx}
\usepackage{booktabs}
\usepackage{subcaption}

\addto\captionsfrench{%
}

\begin{document}

\noindent Preprint submitted in: \\
ArXiv url: \\[0.5cm]

\begin{center}
    \textbf{\Large A high-order ghost-point immersed boundary method with auxiliary differential equations for time-domain impedance walls}\\[0.5cm]
    \textit{A. Ouattara, D. Marx, C. Prax}\\[0.5cm]   
    {Institut PPRIME, CNRS-Université de Poitiers-ENSMA, Poitiers, France}
\end{center}

\textbf{Abstract} This paper presents a method for performing time-domain simulations of a sound field interacting with a wall having a locally-reacting impedance. The linearized Euler equations are solved using a high-order finite difference method. To impose the impedance condition on a wall that is not aligned with the Cartesian grid, an immersed boundary method (IBM) is employed. It relies on a ghost-point approach with a normal-stencil-based reconstruction, where the impedance is accounted for via auxiliary differential equations (ADEs), which are integrated using the same Runge-Kutta scheme as the governing equations. The method is validated against semi-analytical solutions for a pressure pulse reflecting off a flat or cylindrical wall with an impedance corresponding to a mass-spring-damper system. The cylindrical case solution is derived in the appendix. The measured convergence order is approximately 4 for a flat wall and just slighlty smaller for a curved wall. Long-term stability is verified by the diffraction of a monopole by a cylinder.

\section{Introduction}
The finite difference method (FDM) is widely used in computational acoustics and aeroacoustics to simulate sound and flow fields in the time domain. Relying on Cartesian grids, it is relatively simple and computationally efficient. Improving the precision of the FDM, by increasing the formal order of accuracy of the schemes or employing low-dissipative and low-dispersive schemes \cite{BogeyBailly2004}, is also straightforward away from boundaries. Close to boundaries, the precision and stability of the boundary schemes can eventually become an issue. When some diffracting objects, possibly covered with a sound absorber, or ground surfaces are included in the computational domain, their impedance must be considered. The geometry of the object itself may not align with the Cartesian grid.

A non-rigid body can be accounted for in two ways: either by modelling its surface as an impedance boundary condition \cite{BOTTELDOOREN1994}, or by calcutating the sound field inside the body \cite{BLUMRICH2002, SALOMONS2002, HEUTSCHI2005}. This paper considers the first option and, more particularly, a locally reacting surface characterized by its impedance $\tilde{Z}(\omega)$ in the frequency domain. To obtain a time domain impedance boundary condition (TDIBC), this must be expressed in the time domain. The impedance, as a linear time-invariant (LTI) filter, can be equivalently described in the frequency domain by its transfer function (the impedance $Z(\omega)$), or in the time domain by its impulse response or via a differential/difference equation \cite{OPPENHEIM_BOOK}. Enforcing the surface impulse response requires calculating a convolution integral\cite{fung2000, WILSON2007, reymen2008}. This can be computationally intensive in terms of both processing time and memory usage and possibly difficult to couple with a multi-step time integration method, such as a Runge-Kutta scheme. It can be more efficient to simulate the differential equation associated with the impedance. Tam and Auriault \cite{TamAuriault1996} implemented a mass-spring-damper (MSD) mechanical system as a set of two first-order differential equations, integrated simultaneously with the acoustic equations governing the fluid. Being defined directly in the time domain and in a physical way, the continuous MSD system is well-posed. Its discrete numerical implementation can be based on primitive\cite{MARX2015} or characteristic\cite{SEBASTIAN2019} variables. The system corresponds to a second-order frequency response and is thus limited in scope. A more general response is required to account for an arbitrary impedance. Since the transfer function of a LTI filter is a rational function, it can be decomposed into a sum of first and second order filters \cite{GUSTAVSEN1999, OPPENHEIM_BOOK}. Each of these can be associated with an ordinary differential equation in the time domain. This idea has been used  by Bin et al\cite{BIN2009} to simulate impedance boundary conditions, or by Dragna et al \cite{DRAGNA2015} to account for porous media. The latter authors referred to this method as the auxiliary differential equations (ADEs) and showed that these can be advanced using the same time integration method as the rest of the simulation, with preservation of the temporal order of accuracy. Troian et al. \cite{TROIAN2017} used the ADE formulation to model a TDIBC in a linearized Euler solver with flow and performed impedance eduction in a lined duct.

The shape of the boundary bearing the impedance must also be considered. In the simplest case, the boundary aligns with the grid coordinates (which may be curvilinear), and the TDBIC can be directly enforced at the boundary nodes since those belong to the computational grid. However, such alignment is not always possible. In these cases, in the first approximation the boundary can be accounted for by a staircase representation  \cite{HEUTSCHI2005}, but this would destroy the accuracy order of a high-order scheme. To better handle a geometry non-aligned with the grid, the immersed boundary method (IBM) is an alternative. This was first introduced by Peskin~\cite{PESKIN1972} to describe a thin membrane immersed in a fluid. A force was distributed among the fluid points in the neighboring of the membrane in order to impose boundary conditions. The IBM has been developed in several directions since then, with the idea of enforcing boundary conditions at points not belonging to the grid \cite{MITTAL_IACCARINO2005}. A subclass called sharp interface methods\cite{MITTAL2003} rely on ghost points rather than on a distributed force. Geometrically, it reconstructs the field inside the solid at a limited number of ghost nodes, thereby allowing a high-order finite-difference stencil to be applied uniformly across the entire domain, including near the immersed boundaries.

In acoustics, a distributed force has been used by Sun et al \cite{SUN2012} to compute the diffraction of a pulse by a rigid cylinder with the linearized Euler equations. The spatial order was less than 2, which is inherent to the forcing method. A sharp interface method for the linerarized Euler equations was introduced by Seo and Mittal \cite{SeoMittal2011} for acoustics in the presence of rigid bodies, which achieved a fourth-order accuracy. A related method was presented by Zhao et al.~\cite{ZHAO2021} to predict noise of aerodynamic origin with the acoustic perturbation equations, while with a lesser accuracy. Fukushima et al\cite{FUKUSHIMA2016} used a related image point method to compute the radiation of sound by an airplane nacelle. Comparatively, few works have adressed the modeling of non rigid walls with the IBM.
The coupling of an IBM with an impedance condition an has been addressed by Bilbao \cite{Bilbao2022JASA,Bilbao2023JASA}, who used a Peskin-type formulation (continuous forcing) combined with a passivity-preserving discretization of the admittance, and analytically show that the classical CFL stability limit is maintained regardless of the imposed impedance. More recently Bocquet et al~\cite{Bocquet2023JTCA} introduced a boundary condition treatment based on the method of characteristics for moving boundaries.

The objective of this work is to present and validate a coupled ADE–IBM method for modeling an impedance wall with a geometry not aligned with the computational grid.  Section~\ref{sec:method} presents the governing equations, namely the two-dimensional (2D) linearized Euler equations and the wall boundary conditions, including the TDIBC based on the ADE. An ADE formulation of the mass-spring-damper system is given as this will be used to validate the computational method. Section~\ref{sec:numerics} presents the numerical method. The baseline numerical schemes are first presented, and the algorithm coupling IBM with ADE-based boundary condition is described. Section~\ref{sec:results} presents the validation of the method through two  two-dimensional configurations: a flat immersed wall and a circular cylinder reflecting an acoustic pulse, a semi-analytical solution for the latter problem being reported in Appendix~\ref{sec:analytical_reflexion}. The scattering of a harmonic monopole is also addressed. The final section presents the conclusions.

%

\section{Governing Equations}
\label{sec:method}

\subsection{Linearized Euler equations}
\label{subsec:LEE}

The linearized Euler equations offer an ideal framework for studying key aspects of acoustic phenomena, with or without mean flow \cite{TAM1993, TamAuriault1996,SALOMONS2002, HEUTSCHI2005, BIN2009, SeoMittal2011, MARX2015}. No mean flow or inhomogenous propagation medium is considered in the following, in which case the governing equations are:
\begin{equation}
  \frac{\partial {\bf q}}{\partial t} + \frac{\partial {\bf E}}{\partial x} + \frac{\partial {\bf F}}{\partial y} = {\bf S},
  \label{eq:LEE}
\end{equation}
where $t$ represents time and $x$ and $y$ are the cartesian coordinates. The vector of primitive variables is given by ${\bf q}=[p \; \rho_a u \; \rho_a v]^T$, where the exponent ${\rm T}$ denotes the transpose. The initial condition at $t=0$ is ${\bf q}_0 = [p_0 \; 0 \; 0]^{\rm T}$.
The vectors ${\bf E}=[c_a^2\rho_a u \; p \; 0]^{\rm T}$ and ${\bf F}=[\rho_ac_a^2 v \; 0 \; p]^{\rm T}$ are the flux vectors in the $x$- and $y$-directions, respectively. The vector ${\bf S} = [s \; 0 \; 0]^{\rm T}$ contains the source term, where $s$ is monopole for example. In the preceding, $\rho_a$ and $c_a$ denote the density and speed of sound of the ambient fluid. The variable $p$ denotes acoustic pressure and the variables $u$ and $v$ are the $x-$ and $y-$components of the acoustic velocity. The equations above are normalised with $c_a$, $\rho_a$ and a relevant length scale, so that $c_a=1$, $\rho_a=1$ in the following. The boundary conditions at the wall of a body, rigid or of given impedance, will be discussed below. At non-wall boundaries, a non-reflecting boundary condition (NRBC) is employed \cite{TAM1996}, which enforces outgoing cylindrical waves.

\subsection{Wall Boundary conditions}
\label{sec:BoundaryConditions}
In this section, we specify the governing equations for a rigid wall or for a wall of given impedance.

\subsubsection{Rigid wall}
\label{sec:rigid-wall}
A rigid wall corresponds to the familiar impermeability condition, that is, to a zero normal velocity at the body's surface:
\begin{equation}
    u_n = 0 .
    \label{eq:rigid-un}
\end{equation}
A companion pressure condition follows from the momentum equation in
Eq.~\eqref{eq:LEE} projected on the wall normal $\mathbf n$:
\begin{equation}
  \rho_a\,\partial u_n/\partial t = -\partial p/\partial n .
  \label{eq:momentum-normal}
\end{equation}
Since Eq.~\eqref{eq:rigid-un} holds at every instant, its time derivative vanishes identically, and Eq.~\eqref{eq:momentum-normal} gives the Neumann condition for the pressure at a rigid wall :
\begin{equation}
\frac{\partial p}{\partial n} = 0.
\label{eq:rigid-dpdn}
\end{equation}
Equations \eqref{eq:rigid-un}--\eqref{eq:rigid-dpdn} are the Dirichlet (velocity) and Neumann (pressure) boundary conditions typically used to design immersed boundary methods (IBM) \cite{MITTAL_IACCARINO2005,SeoMittal2011,BREHM2015}. The method on which the present paper is based, presented in Section~\ref{sec:immersedboundary}, was originally developed for rigid walls \cite{Diaz2022CFA,Ouattara2025These,ouattara2026IBM}.


\subsubsection{Time domain impedance boundary conditions with ADE}
\label{sec:ade-general}
A locally-reacting boundary is a LTI system that relates the acoustic pressure $p$ at a point on the surface to the normal velocity $u_n = \mathbf u\cdot\mathbf n$ at that point, where $\mathbf n$ is the surface normal pointing from the fluid into the solid. The local nature of the impedance means the response at a point is independent of both the angle of incidence and the spatial distribution of the sound field. This is particularly appropriate for the single and double degree of freedom resonant liners used for example in aeronautics\cite{TROIAN2017}. 
The impedance is the transfer function of the LTI system, $Z(s)$, expressed as:
\begin{equation}
    \label{eq:def_impedance}
    \hat p(s) = Z(s)\,\hat u_n(s)
\end{equation}
where $\hat{\phantom{p}}$ denotes the one-sided Laplace transform (LT), which for a function $f(t)$ is given by:
\begin{equation}
    \label{eq:LT}
    \hat{f}(s) = \int_0^\infty f(t) e^{-s t} dt
\end{equation}
Given the normalization in Section~\ref{subsec:LEE}, $Z(s)$ is normalized by $\rho_a c_a$. Sometimes, it is more convenient to use the admittance $Y(s) = 1/Z(s)$ verifying:
\begin{equation}
    \label{eq:def_admittance}
    \hat u_n(s) = Y(s)\hat p(s) \,
\end{equation}
The complex Laplace variable is expressed as $s = \alpha + i\omega$. Often, the impedance is given as the frequency response $\tilde{Z}(\omega) = Z(s = i\omega)$, i.e., the transfer function restricted to the imaginary axis. The full transfer function provides deeper insight into system properties such as stability, which is determined by the location of the poles.

Now, a time-domain implementation of Eq.~(\ref{eq:def_impedance}) is needed. One option would be to take the inverse Laplace transform of Eq.~(\ref{eq:def_impedance}) to form a convolution integral; this has been discarded for reasons explained in the introduction.
The present work uses the ADE method\cite{DRAGNA2015,TROIAN2017}, which turns Eq.~(\ref{eq:def_impedance}) into a small, fixed number of ordinary differential boundary equations that can be incorporated into Eq.~(\ref{eq:LEE}). The ADE method starts from the transfer function written as a rational function. This is directly available if the differential equation of the system is known, or it can be obtained by fitting the experimentally obtained frequency response  \cite{GUSTAVSEN1999}. A partial fraction decomposition of the transfer function leads to the following expression, here written for the admittance\cite{DRAGNA2015}:
\begin{equation}
    Y(s) = Y_\infty + \sum_{k=1}^{N} \frac{A_k}{s-\lambda_k}
    + \sum_{k=1}^{M} \left(\frac{q_k}{s-s_k} + \frac{q_k^{*}}{s-s_k^{*}}\right)
     \label{eq:Yrational}
\end{equation}
where $^*$ indicates the complex conjugate, $\lambda_k$ are $N$ simple real poles, and $(s_k, s_k^*)$ are $M$ complex-conjugate pole pairs, with $s_k = \alpha_k + i \omega_k$. The denominators $A_k$ are real and the denominators $q_k$ are complex, with $q_k = B_k+iC_k$.

The inverse LT of Eq.~(\ref{eq:Yrational}) provides the impulse response function of the system, which can be injected into the convolution equation resulting from the inverse LT of Eq.~(\ref{eq:def_admittance}). After some algebra, this gives the following  expression for the wall normal velocity\cite{DRAGNA2015}:
\begin{equation}
\begin{split}
    u_n(t) = & Y_\infty p(t) + \sum_{k=1}^{N} A_k\Phi_k(t) \\
             &+ \sum_{k=1}^{M} 2\bigl[B_k\Psi_k^{(1)}(t) - C_k\Psi_k^{(2)}(t)\bigr] ,
             \end{split}
\label{eq:unGeneral}
\end{equation}
where $\Phi_k(t)$, $\Psi_k^{(1)}(t)$, and $\Psi_k^{(2)}(t)$ are auxiliary functions governed by the ADE:
\begin{subequations}
  \label{eq:ADEgeneral}
  \begin{align}
  \label{eq:ADEgeneral_a}
  \dfrac{d\Phi_k}{dt} = & \lambda_k\Phi_k + p, \qquad \\
  \label{eq:ADEgeneral_b}
  \dfrac{d\Psi_k^{(1)}}{dt} = & \alpha_k\Psi_k^{(1)} - \omega_k\Psi_k^{(2)} + p, \qquad \\
  \label{eq:ADEgeneral_c}
  \dfrac{d\Psi_k^{(2)}}{dt} = & \alpha_k\Psi_k^{(2)} + \omega_k\Psi_k^{(1)} .
  \end{align}
\end{subequations}
For any point of the surface, there  is one real auxiliary function $\Phi_k$ per simple pole and two real auxiliary functions $\Psi_k^{(1)},\Psi_k^{(2)}$ per complex-conjugate
pair, all forced by the pressure $p$ at that point. Given the acoustic pressure, the ADE method consists in integrating the differential equations in Eq. (\ref{eq:ADEgeneral}) and injecting the resulting auxiliary functions into Eq. (\ref{eq:unGeneral}), which provides the normal velocity, thereby implementing the TDIBC. Equation (\ref{eq:unGeneral}) is the counterpart to the rigid-wall equation, Eq. (\ref{eq:rigid-un}).

\subsubsection{Single-resonance mass spring damper wall}
\label{sec:impedance-msd}

For the validation of the numerical method hereafter, a MSD system is used. This second order system was first employed by Tam and Auriault \cite{TamAuriault1996} who presented their own set of equations for a TDIBC. The objective here is to write this system in a form similar to Eq.~(\ref{eq:Yrational}) so that the ADE can be used in a systematic way.

Given a locally reacting wall corresponding to mass $M_m$, damping $R_m$, and stiffness $K_m$ (all real positive), its impedance is $Z(s) = R_m + M_ms + \frac{K_m}{s}$, which corresponds to an admittance:
\begin{equation}
    \label{eq:MSD_admittance}
    Y(s) = \frac{1}{Z(s)} = \frac{s/M_m}{(s-s_{+})(s-s_{-})}
\end{equation}
with $s_\pm$ the roots of $M_ms^2+R_ms+K_m=0$,
\begin{equation}
s_\pm = -\frac{R_m}{2M_m} \pm \sqrt{\left(\frac{R_m}{2M_m}\right)^2 - \frac{K_m}{M_m}}.
\end{equation}
For $M_m,R_m,K_m>0$ both roots lie in the open left half-plane, so the wall
is causal and passive. 
In addition, the two poles are real when $K_m/M_m < (R_m/2M_m)^2$ and form a complex-conjugate pair when $K_m/M_m > (R_m/2M_m)^2$. Only this latter situation will be considered and corresponds to Eq.~(\ref{eq:Yrational}) with: $Y_\infty=0$, $N=0$, $M=1$, $q_1 = s_+/[M_m(s_+-s_-)]$, $\alpha_1 = -\frac{R_m}{2M_m}$,  and $\omega_1=\sqrt{\frac{K_m}{M_m}-\left(\frac{R_m}{2M_m}\right)^2}$. For a pair of complex conjugate poles, Eq.~\eqref{eq:ADEgeneral} reduces to a pair of equations for the auxiliary states $\Psi^{(1)}$ and $\Psi^{(2)}$,
\begin{equation}
    \label{eq:ADE_AUXEQ_MSD}
    \begin{split}
    \dfrac{d\Psi^{(1)}}{dt} = \alpha_1\Psi^{(1)} - \omega_1\Psi^{(2)} + p,\\
    \dfrac{d\Psi^{(2)}}{dt} = \alpha_1\Psi^{(2)} + \omega_1\Psi^{(1)}.
    \end{split}
\end{equation}
The wall normal velocity is then given by Eq.~(\ref{eq:unGeneral}) which reduces to:
\begin{equation}
    u_{n} = 2B_1\Psi^{(1)} - 2C_1\Psi^{(2)},
    \label{eq:ADE-un}
\end{equation}
where $B_1$ and $C_1$ are the real and imaginary parts of $q_1$.
The time derivative of $u_n$ will be needed for the numerical implementation and is given by: 
\begin{equation}
\dfrac{d u_{n}}{d t} = 2B_1\dfrac{d\Psi^{(1)}}{dt} - 2C_1\dfrac{d\Psi^{(2)}}{dt}.
\label{eq:ADE-dundt}
\end{equation}
where the derivatives are substituted from Eq.~(\ref{eq:ADE_AUXEQ_MSD}). Equations~ (\ref{eq:ADE_AUXEQ_MSD}-\ref{eq:ADE-dundt}) are the one to be imposed as a TDIBC based on ADE for a MSD system.

\section{Numerical method}
\label{sec:numerics}
In this section, the baseline numerical schemes are first presented. This is followed by the presentation of the algorithm coupling the IBM with the ADE-based boundary condition.

\subsection{Numerical schemes}
\label{sec:numerical_schemes}
The grid is Cartesian, with grid nodes located at coordinates $x_i$ and $y_j$, $i=1...n_x, \; j=1...n_y$. The mesh sizes in the $x$ and $y$-directions are equal, uniform in space, and denoted by $h$. All variables are collocated at nodes $(x_i, y_j)$, with ${\bf q}_{ij} = [p_{ij} \; \rho_a u_{ij} \; \rho_a v_{ij}]^{\rm T}$. The governing equation (\ref{eq:LEE}) can be rewritten as :
\begin{equation}
  \dfrac{\partial {\bf q}_{ij}}{\partial t} = \text{RHS}_{ij},
  \label{eq:EDO}
\end{equation}
where $\text{RHS}_{ij}$ denotes the right-hand side of the governing equations, which depends on ${\bf q}$ on the grid. To advance such a system in time, a 4th order 4-stage Runge-Kutta method  is used\cite{MITCHELL1995}. The numerical spatial derivatives in $\text{RHS}_{ij}$ are computed with the DRP scheme \cite{TAM1993,TAM1994}. The first $x$-derivative of a variable $f$ at the point $i$ is given by:
\begin{equation}
    \left.\dfrac{\partial f}{\partial x}\right)_i = \dfrac{1}{h} \sum_{k=-3}^{3} a_k f_{i+k},
    \label{DRP_s}
\end{equation}
This is a seven-point stencil that could lead to a formal order of accuracy up to 6. However, the DRP scheme is only 4th-order, using extra degrees of freedom to be less dispersive and dissipative compared to regular 4th-order finite differences. It is thus well suited for wave propagation problems and aeroacoustics. This scheme is used in the interior of the fluid, and also close the solid boundaries where ghost points can be used, as detailed below. For the external boundaries of the domain at which NRBC are applied, a 4th order one-sided DRP schemes is used. The coefficients for all these schemes are given by Tam and Dong\cite{TAM1994}.\\

\subsection{Immersed boundary method}
\label{sec:immersedboundary}



The equations to be satisfied at a rigid wall are Eq.~(\ref{eq:rigid-un}) and possibly Eq.~(\ref{eq:rigid-dpdn}). For a TDIBC based on ADE, Eqs.~(\ref{eq:unGeneral}-\ref{eq:ADEgeneral}), or Eqs.~(\ref{eq:ADE_AUXEQ_MSD}-\ref{eq:ADE-un}) for a MSD system, should be imposed. These can be advanced in time with the same Runge-Kutta method as for the interior equations, that is, with the same order of accuracy. However, for a boundary non-aligned with the computational grid, these equations should be enforced at boundary points not belonging to the grid. For this purpose, an IBM is used in the present work, which follows the ghost-point approach developed by Diaz et al. \cite{Diaz2022CFA}.


\subsubsection{Ghost-point reconstruction}
\label{sec:ibm}
\begin{figure}[ht!]
\centering
\includegraphics[width=0.6\linewidth]{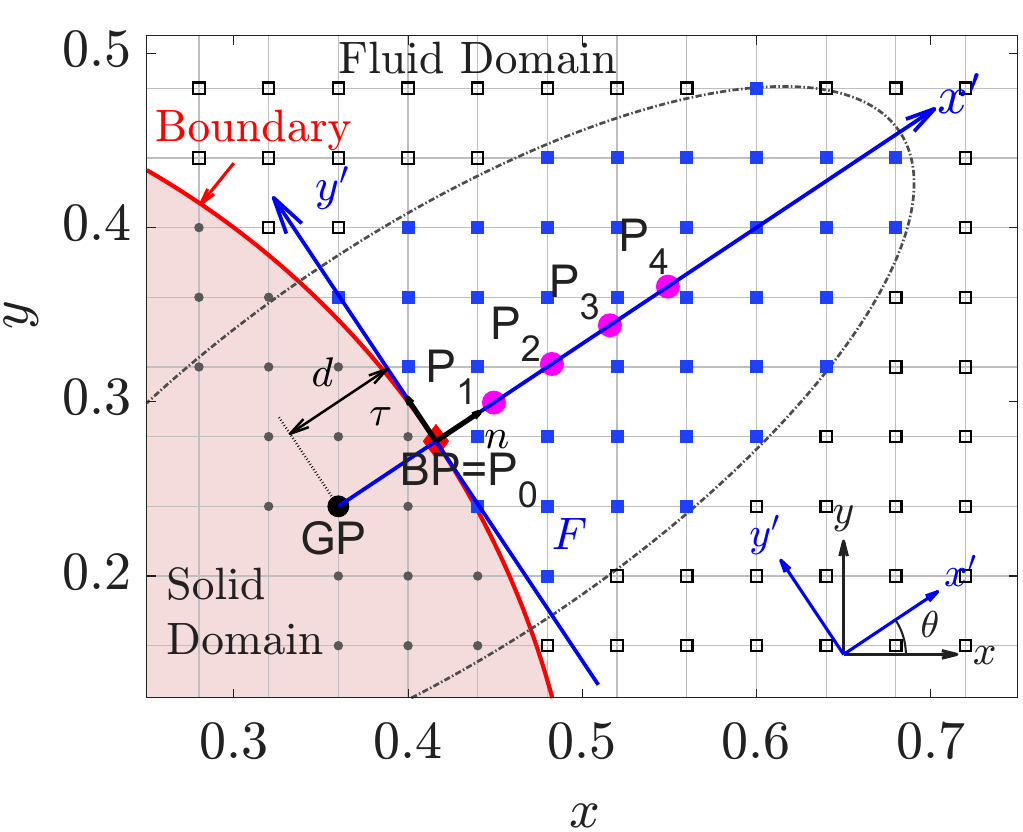}
\caption{Cartesian grid in the vicinity of a curved boundary separating fluid and solid (exterior) regions, showing : a particular ghost point (GP), its associated Normal-stencil (points $P_i$) including the boundary point (BP, which is also $P_0$). Fluid nodes are indicated by squares, those participating to the 2D least-squares fit belong to the elliptical cloud and are shown by filled square symbols. The $x'$-axis is aligned with the local wall normal vector $\mathbf{n}$ at the BP, which makes an angle $\theta$ with the $x$-axis. The tangent vector at the BP is $\boldsymbol{\tau}$.}
\label{fig:schema}
\end{figure}

Referring to Fig.\ref{fig:schema}, the IBM begins by identifying the Cartesian grid points belonging to the fluid and solid domains, using a mask function. Equation(\ref{eq:EDO}) is solved only for the nodes located in the fluid domain. Fluid and solid points are separated by the interface where the boundary condition must be applied.
%
The boundary is responsible for two related problems. First, considering a fluid node adjacent to the boundary (point F for example), the spatial derivation scheme in Eq.~(\ref{DRP_s}) cannot be applied, here in the $x$-direction, since this would require three neighbors on the left of point F that are not available, being located within the solid. Second, the boundary condition itself should be imposed at points of the boundary, but most of them do not belong to the grid. The ghost points (GP) are fictitious points that are created within the solid in the vicinity of the boundary to provide the missing neighbors so that the numerical scheme can be applied at any point in the fluid, and in addition the values of the variables ($u$, $v$, $p$) at these GP are specified in such a way that the boundary condition is true at the very location of the boundary.
For the 7-point stencil presented in Eq.~(\ref{DRP_s}) three layers of GPs in the solid adjacent to the boundary are necessary, as shown in Fig.~(\ref{fig:schema}). Each GP is associated with a boundary point (BP), which is its normal projection on the boundary, and the distance from the GP to the BP is denoted by $d$. The determination of the BP  (also called the boundary intercept point) given a GP has been discussed in the literature\cite{MITTAL2003, SeoMittal2011}.
At the BP, a local frame $(\boldsymbol{n}, \boldsymbol{\tau})$ is defined, where $\boldsymbol{n}=(n_x,n_y)$ is the unit normal vector (here, pointing into the fluid), and $\boldsymbol\tau$ is the unit tangent vector. In this frame, the coordinates are denoted by $x'$ and $y'$, which are centered at BP and scaled by the uniform grid spacing $h$. These are defined by:
\begin{equation}
\begin{split}
\begin{pmatrix}x'\\ y'\end{pmatrix}
= T\,\frac{1}{h}\begin{pmatrix}x-x_{\mathrm{BP}}\\ y-y_{\mathrm{BP}}\end{pmatrix},
\qquad \\
T = \begin{pmatrix} n_x & n_y \\ -n_y & n_x \end{pmatrix}
= \begin{pmatrix}\cos\theta & \sin\theta \\ -\sin\theta & \cos\theta\end{pmatrix},
\label{eq:ns-rotation}
\end{split}
\end{equation}
where $T$ is a rotation of angle $\theta$, the angle between $\mathbf n$ and the $x$-axis (see Fig.~\ref{fig:schema}). This rotation is also employed to compute the normal and tangential components of velocity, $(u_n, u_t)$, from the components $(u,v)$. 

To enforce the value of the fields at a given GP, a 1D extrapolation along the normal direction is used. This is based on points $P_0, P_1, ..., P_m$ evenly distributed with a step equal to $h$ (the mesh size) along the normal, see Fig.~\ref{fig:schema}. Point $P_0$ coincides with BP. The stencil of points $\{P_0, ...,P_m\}$ is referred to as the normal stencil \cite{Diaz2022CFA}. Since the points $P_i$ do not belong to the grid, the values of the fields at these points need to be obtained from the surrounding fluid points by interpolation, using a 2D polynomial least-squares. Therefore, the strategy consists in two steps: a 2D least-squares fit is performed to obtain the variables at the points of the normal stencil, which is followed by a 1D extrapolation along the normal to obtain the value at the GP. The order of each step is chosen to achieve the desired accuracy, as the IBM must enforce GP values without significantly degrading the global order of the numerical method, which is 4 in this case. This two-step procedure was originally designed for use with characteristic variables \cite{Diaz2022CFA}; however, it is here applied to primitive variables. Some more details are now given about the two interpolation steps.

\paragraph{Step 1: 2D least-square fit}
For each BP, a cloud of fluid points is first defined to construct a 2D interpolating polynomial of order $k$ and obtain the field values at points $P_i$. To be consistent with the order of accuracy of the other numerical schemes, $k=4$ is taken.
The points of the cloud are located within an ellipse, centered at the BP, in the region defined by:
\begin{equation}
    \varepsilon = \left\{ (x',y')\in\Omega_f\ ;\ \left(\frac{x'}{R_n}\right)^{\!2} + \left(\frac{y'}{R_t}\right)^{\!2} \leq 1 \right\}
    \label{eq:ns-ellipse}
\end{equation}
where $\Omega_f$ denotes the fluid domain, $R_n= 2k$ and $R_t= \frac{k}{2}+1$ are the ellipse half-widths in the normal and tangential directions, respectively. The ellipse is chosen to be wide enough along the normal direction and narrow enough along the tangential direction to reach the farthest normal stencil point,  $P_m$. For every field $f\in\{u_n,u_t,p\}$, with values $f_{ij}$ at the grid points, the $k$-th order 2D interpolating polynomial is:
\begin{equation}
\mathcal{P}(x',y') = \sum_{\alpha=0}^{k}\ \sum_{\beta=0}^{k} a_{\alpha\beta}\,(x')^{\,\alpha}(y')^{\,\beta} , \quad \alpha+\beta \le k
\label{eq:ns-poly2d}
\end{equation}
The constant coefficients $a_{\alpha\beta}$ are associated with the monomials $(x')^{\,\alpha}(y')^{\,\beta}$. The number of monomials, $M=(k+1)(k+2)/2$, and the number of data points in the ellipse, $N>M$, are such that finding the coefficients $a_{\alpha\beta}$ is an over-determined problem, set by imposing that:
\begin{equation}
\mathcal{P}(x'_i,y'_j) = f_{ij}, \qquad \forall (x_i',y_j')\in \varepsilon.
\label{eq:ns-collocation}
\end{equation}
Equations (\ref{eq:ns-poly2d}-\ref{eq:ns-collocation}) is written as the over-determined linear system $\mathbf f = V\mathbf a$, with
\begin{equation}
V = \begin{pmatrix}
1 & x'_1 & y'_1 & \cdots & x_1'^{\,k} & y_1'^{\,k} \\
1 & x'_2 & y'_2 & \cdots & x_2'^{\,k} & y_2'^{\,k} \\
\vdots & \vdots & \vdots & & \vdots & \vdots \\
1 & x'_N & y'_N & \cdots & x_N'^{\,k} & y_N'^{\,k}
\end{pmatrix},
\qquad
\mathbf a = \begin{pmatrix}a_{00}\\ a_{10}\\ a_{01}\\ \vdots\end{pmatrix},
\label{eq:ns-vandermonde}
\end{equation}
where $V$ is a Vandermonde matrix. This system is solved in the least-squares sense, with a formal solution given by :
\begin{equation}
\mathbf a = \left(V^{\top}V\right)^{-1}V^{\top}\,\mathbf f.
\label{eq:ns-LSQ}
\end{equation}

Finally, once the coefficients are known, the polynomial of Eq.~\eqref{eq:ns-poly2d} is evaluated at points $P_i$, providing the field values at those points:
\begin{equation}
  f(P_i) = \mathcal{P}(x'_i,y'_i),  \quad\forall i=0,\dots,m .
  \label{eq:ns-readout}
\end{equation}
This is the final outcome of the first interpolation.\\

\paragraph{Step 2: Constrained extrapolation along the normal direction}
A one-dimensional polynomial of degree $k=4$ is then used to extrapolate along the normal direction the fields $f(P_i)$ of step 1 to the ghost points. This polynomial is:
\begin{equation}
    \mathcal{Q}(x')=\sum_{j=0}^{k} b_j\,x'{}^j , 
\end{equation}
where, in particular, $b_0=Q(0)$ and $b_1 = Q'(0)$, with $Q'$ the derivative of $Q$. At this stage no least-squares fit is needed (one was already performed in the previous step); therefore, 5 points $P_i$ suffice to determine the degree-4 polynomial, i.e. $m = 4$. To determine the coefficients $b_j$, $Q$ is evaluated at each normal stencil point $P_i$. The normalized normal coordinate of these points is denoted by $x'_i$. Given the definition of $P_i$ and Eq.~(\ref{eq:ns-rotation}) for $x'$, $x'_i = i$ for all $i = 0, \ldots, m$. In particular, point $P_0$ (which is also BP) corresponds to $x'_0 = 0$. The polynomial evaluated at $x_i'$, $Q(x'_i)$, is nothing but $f(P_i)$ for the variable $f$ under consideration. Writing $\mathcal{Q}(x_i')=\sum_{j=0}^{k} b_j\,x_i'{}^j = f(P_i)$ for $i=0, ..., m$, leads to the following system (with $k=m=4$):
\begin{equation}
    \begin{pmatrix}
        x_0' {}^0 & x_0' {}^1 & x_0' {}^2 & x_0' {}^3 & x_0' {}^4 \\
        x_1' {}^0 & x_1' {}^1 & x_1' {}^2 & x_1' {}^3 & x_1' {}^4 \\
        x_2' {}^0 & x_2' {}^1 & x_2' {}^2 & x_2' {}^3 & x_2' {}^4 \\
        x_3' {}^0 & x_3' {}^1 & x_3' {}^2 & x_3' {}^3 & x_3' {}^4 \\
        x_4' {}^0 & x_4' {}^1 & x_4' {}^2 & x_4' {}^3 & x_4' {}^4 
    \end{pmatrix} 
    \begin{pmatrix}
        b_0\\
        b_1\\
        b_2\\
        b_3\\ 
        b_4
    \end{pmatrix} =
    \begin{pmatrix}
        f(P_0)\\
        f(P_1)\\
        f(P_2)\\
        f(P_3)\\ 
        f(P_4)
    \end{pmatrix}
    \label{eq:syst_extrapol_general}
\end{equation}
The way this system is solved depends on the type of boundary condition at the BP. For a Dirichlet boundary condition $f(P_0)$ is imposed (e. g., for the normal velocity). For a Neumann boundary condition, $f'(P_0)$ is imposed (e.g., for the pressure), where $f' = \partial f / \partial x'$ is the normal derivative. If no boundary condition exists (e.g., for the tangential velocity), a Dirichlet boundary condition is used but $f(P_0))$ results from the interpolation in step 1 rather than from a boundary condition; the ghost point value is then enforced by free extrapolation.

For a Dirichlet condition, $f(P_0)$ is given, and $f(P_i), \, i>0$ results from the interpolation of step 1. Therefore, Eq.~(\ref{eq:syst_extrapol_general}) is used directly, yielding an invertible $5 \times 5$ system whose solution provides the coefficients $b_i$. Once these are known, the value at the ghost point is obtained by evaluating:
\begin{equation}
    f(\mathrm{GP}) = Q(x'_{\mathrm{gp}}) = \sum_{j=0}^{k} b_j x'_{\mathrm{gp}}{}^j, 
    \label{eq:f_at_GP}
\end{equation}
where $x'_{\mathrm{gp}} = -d/h$ is the normalized coordinate of the GP along the normal direction.

For a Neumann condition, the derivative $f'(P_0)$ of the field (which equals $b_1$) is known, while $f(P_0)$ (which equals $b_0$) is unknown. Therefore, in Eq.~(\ref{eq:syst_extrapol_general}), the known term involving $b_1$ is moved to the right-hand side, and the first row of the system is removed. This yields:
\begin{equation}
    \begin{pmatrix}
        x_1' {}^0 & x_1' {}^2 & x_1' {}^3 & x_1' {}^4 \\
        x_2' {}^0 & x_2' {}^2 & x_2' {}^3 & x_2' {}^4 \\
        x_3' {}^0 & x_3' {}^2 & x_3' {}^3 & x_3' {}^4 \\
        x_4' {}^0 & x_4' {}^2 & x_4' {}^3 & x_4' {}^4 
    \end{pmatrix} 
    \begin{pmatrix}
        b_0\\
        b_2\\
        b_3\\ 
        b_4
    \end{pmatrix} =
    \begin{pmatrix}
        f(P_1)\\
        f(P_2)\\
        f(P_3)\\ 
        f(P_4)
    \end{pmatrix}
    + f'(P_0)
    \begin{pmatrix}
        x_1'\\
        x_2'\\        
        x_3'\\    
        x_4'
    \end{pmatrix}
    \label{eq:syst_extrapol_neumann}
\end{equation}
The right-hand side is fully known, and the $4 \times 4$ system on the left-hand side is invertible. Inversion yields $b_0$, $b_2$, $b_3$, and $b_4$, and $b_1$ is known. The field value at GP results again from Eq.~(\ref{eq:f_at_GP}).\\

Once the fields are known at all GPs, the numerical derivatives and hence $\text{RHS}_{ij}$ in Eq.~(\ref{eq:EDO}) can be evaluated at any fluid point, while accounting for the boundary.

\subsubsection{Coupling the immersed boundary reconstruction with the ADE}
\label{sec:coupling}

The IBM was detailed in the previous section, where it was shown that the boundary condition is imposed during the 1D extrapolation along the normal direction, enforcing the fields at the GPs. This requires the values of $f(P_0)$ or $f'(P_0)$ when these quantities are constrained by the boundary condition. In this section, we detail how these quantities are updated using the ADE during the time-stepping. At each of the four Runge-Kutta stages, the impedance wall and the immersed boundary reconstruction are coupled through the following six steps, taken for each GP and its associated BP:
\\
\indent 1. The pressure at any BP, $p_{\mathrm{BP}}$, is computed from the fluid points using the 2D least-squares fit, as it serves as the forcing term for the ADE evaluated in the next step. Similarly, the tangential velocity at the BP, $u_{t,\mathrm{BP}}$, is computed, as it is required for applying the boundary condition. Here, $p_{\mathrm{BP}}$ and $u_{t,\mathrm{BP}}$ correspond to $f(P_0)$ used above when $f$ represents $p$ and $u_t$, respectively.\\
\indent 2. Using $p_{\mathrm{BP}}$ as the forcing term in the ADE, either Eq.~(\ref{eq:ADEgeneral}) or its specific form for a MSD system, Eq.~(\ref{eq:ADE_AUXEQ_MSD}), these equations are advanced by one Runge-Kutta stage to compute the auxiliary states $\Phi_k$ and $\Psi_k^{(1,2)}$, which must be defined for every BP.\\
\indent 3. The wall-normal velocity, $u_{n,\mathrm{BP}}$, is then computed at every BP using either Eq.~(\ref{eq:unGeneral}) or its specific form for a MSD, Eq.~(\ref{eq:ADE-un}). Its time derivative is also computed by taking the time derivative of Eq.~(\ref{eq:unGeneral}), as shown in Eq.~(\ref{eq:ADE-dundt}) for a MSD.\\
\indent 4. The boundary condition is accounted for as follows. A Dirichlet boundary condition is applied to the normal velocity, which is set to $u_{n,\mathrm{BP}}$ from step 3. For pressure, a Neumann condition is applied, with its derivative derived from the momentum equation projected onto the normal direction: $\partial p / \partial n = -d u_{n,\mathrm{BP}} / dt$, where $d u_{n,\mathrm{BP}} / d t$ is also obtained from step 3. Since the tangential velocity is not constrained, a free reconstruction is used, meaning a Dirichlet condition is applied to $u_t$, which should take the value $u_{t,\textrm{BP}}$ computed in step 1. Note that for a rigid wall, $u_n = \partial p / \partial n = 0$, so steps 2–3 above are simply skipped.\\
\indent 5. These Dirichlet (velocity) and Neumann (pressure) conditions at the BP supply the value of the function or its derivative needed at $P_0$ for the 1D Lagrange extrapolation in the normal direction. The extrapolation determines the values of all the fields at the GPs, by imposing Eq.~(\ref{eq:f_at_GP}).\\
\indent 6. The fields being known at the GPs, the DRP derivatives and $\text{RHS}_{ij}$ in Eq.~(\ref{eq:EDO}) are computed for all fluid points, allowing for the time advancement of the full state column vector $\mathbf{Q} = [\mathbf{p};\mathbf{u};\mathbf{v};\mathbf{\Phi_k};\mathbf{\Psi_k^{(1)}};\mathbf{\Psi_k^{(2)}}]$ which reduces to $\mathbf{Q} = [\mathbf{p};\mathbf{u};\mathbf{v};\mathbf{\Psi^{(1)}};\mathbf{\Psi^{(2)}}]$ for a MSD. The next Runge-Kutta stage can be initiated.\\

Overall, one 2D least-squares interpolation followed by a 1D extrapolation, and one ADE evaluation per Runge-Kutta stage and per ghost point is required to impose the TDBIC.

\section{Results}
\label{sec:results}
To test the present ADE-IBM coupling, three 2D acoustic scattering problems are considered. For each case, the simulation is validated against a semi-analytical solution. Convergence is verified.

\subsection{Flat immersed wall}
\label{sec:flat}
\begin{figure}[ht!]
    \centering
    \begin{subfigure}[t]{0.5\textwidth}
        \centering
        \includegraphics[width=\linewidth]{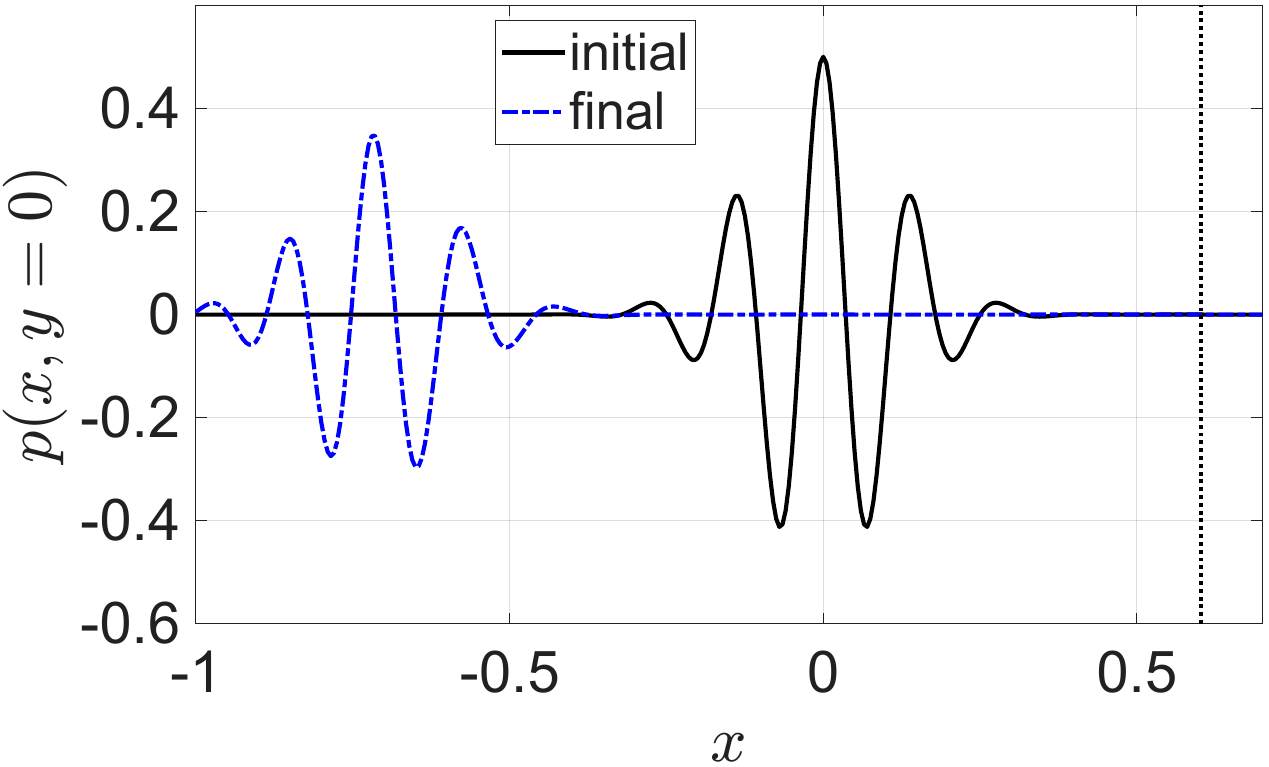}
        \caption{}
    \end{subfigure}%
    ~ 
    \begin{subfigure}[t]{0.5\textwidth}
        \centering
        \includegraphics[width=\linewidth]{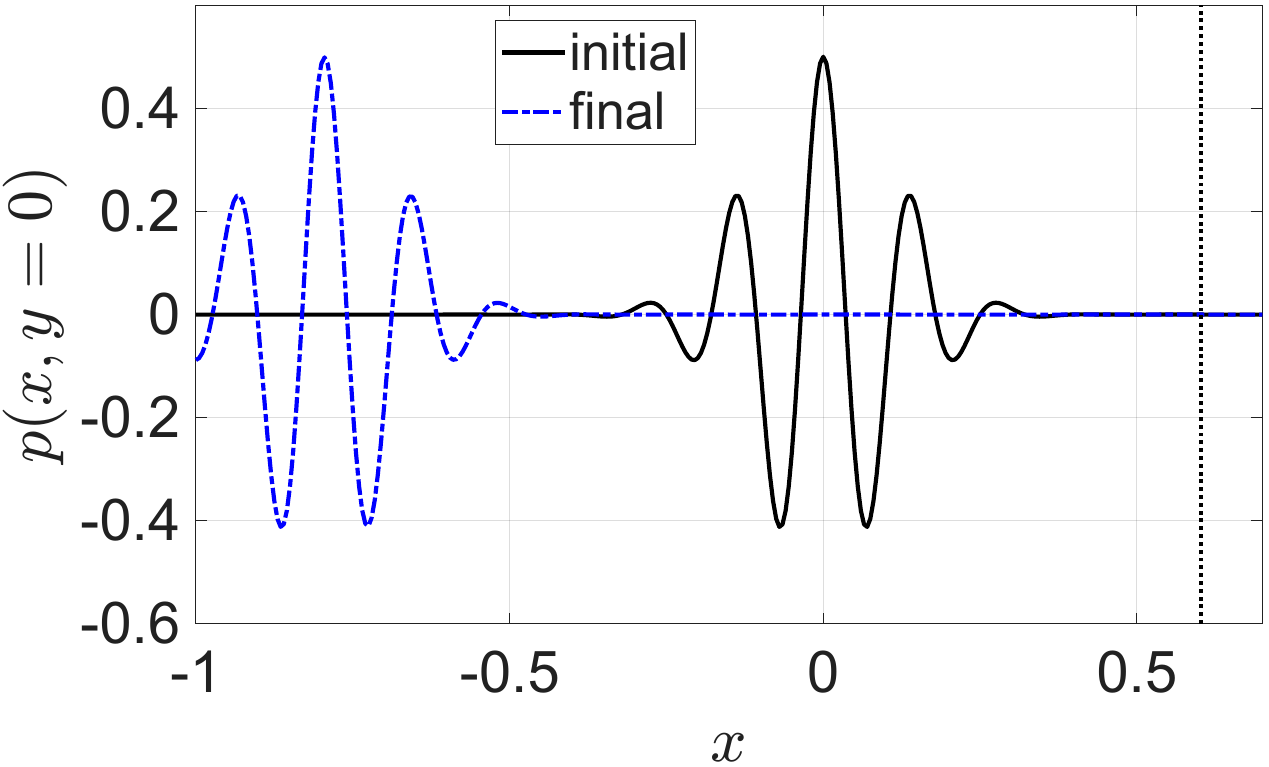}
        \caption{}
    \end{subfigure}
    \caption{Pressure field at initial and final times along the axis $x$ for $y=0$ : (a) impedance wall; (b) rigid wall. The dotted vertical line indicates the wall position.}
    \label{fig:champ_init_fin}
\end{figure}

The first benchmark is a plane wave packet, modulated at the resonance of the wall, reflected by a flat wall immersed at a normal angle to the Cartesian grid. A periodic boundary conditions are imposed in the $y$-direction, therefore the configuration is quasi-one-dimensional, and a semi-analytical solution is available\cite{TamAuriault1996}.\\
The computational domain is $[-1,0.8]\times[-0.2,0.2]$, discretized with a uniform Cartesian grid, with a mesh size $h = 0.005$ and the time step is $\Delta t = 0.0025$. The initial wave packet is given by
\begin{equation}
    \begin{split}
    p_0(x,y) = \frac{a}{2}\, e^{-\alpha^2 (x-x_0)^2}\, \cos\!\big(2\pi k (x-x_0)\big),\\
    u_0 = p_0, \quad v_0 = 0
    \end{split}
\end{equation}
whith $x_0=0$, $a=1$, $\alpha=k/\sqrt(0.8)$ and $k=7$.
With $u_0=p_0$, the wave packet initially propagates in the positive $x$-direction.
The immersed wall is located at $x_\mathrm{wall}=0.6+h/2$, between two grid points, 
so that the wall is not aligned with the grid. 
The grid points to the right of the wall are masked as solid points, and the solver only advances the fluid domain to the left of the wall. 3 layers inside the solid are used as GPs, and the fields at these GPs are enforced with the IBM method presented in Sec.~\ref{sec:immersedboundary}. The NRBC is imposed at the left boundary, $x=-1$, so that the wave packet reflected by the wall can leave the domain without reflection.
Both rigid and impedance walls are considered. The impedance follows Eq.~(\ref{eq:MSD_admittance}) with $M_m=0.025$, $K_m=40$, $R_m=0.2$. The resonance frequency $\omega_{\rm res} = \sqrt{K_m/M_m}$ approximately matches the center frequency ($2\pi k$) excited by the pulse.
Figure~\ref{fig:champ_init_fin} shows the pressure field at the initial time, $t=0$, and at the final time, $t=t_\mathrm{end} = 2$, after reflexion by the wall.
%
The energy of the wave packet is fully reflected by the rigid wall while it is partially absorbed by the impedance wall, which is the expected behaviour.
%
\begin{figure}[ht!]
    \centering
    \begin{subfigure}[t]{0.5\textwidth}
        \centering
        \includegraphics[width=\linewidth]{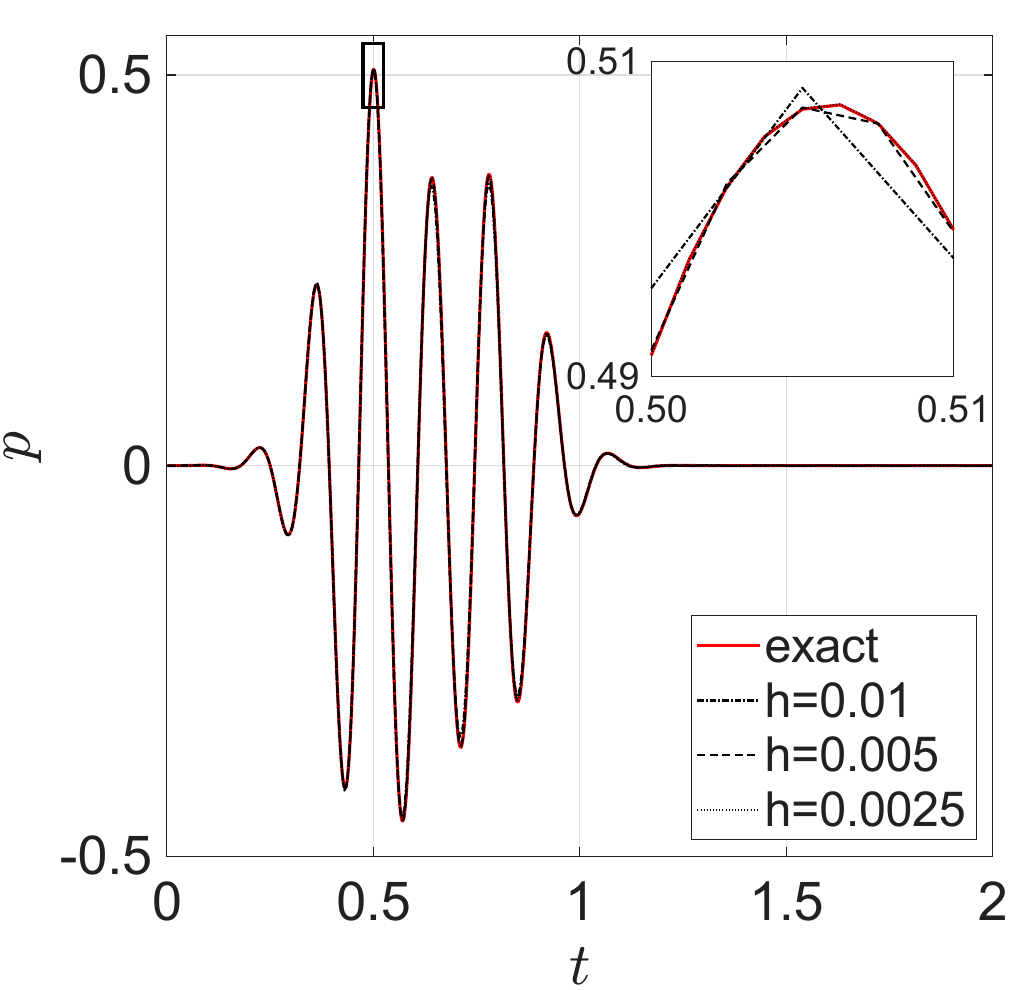}
        \caption{}
    \end{subfigure}%
    ~ 
    \begin{subfigure}[t]{0.5\textwidth}
        \centering
        \includegraphics[width=\linewidth]{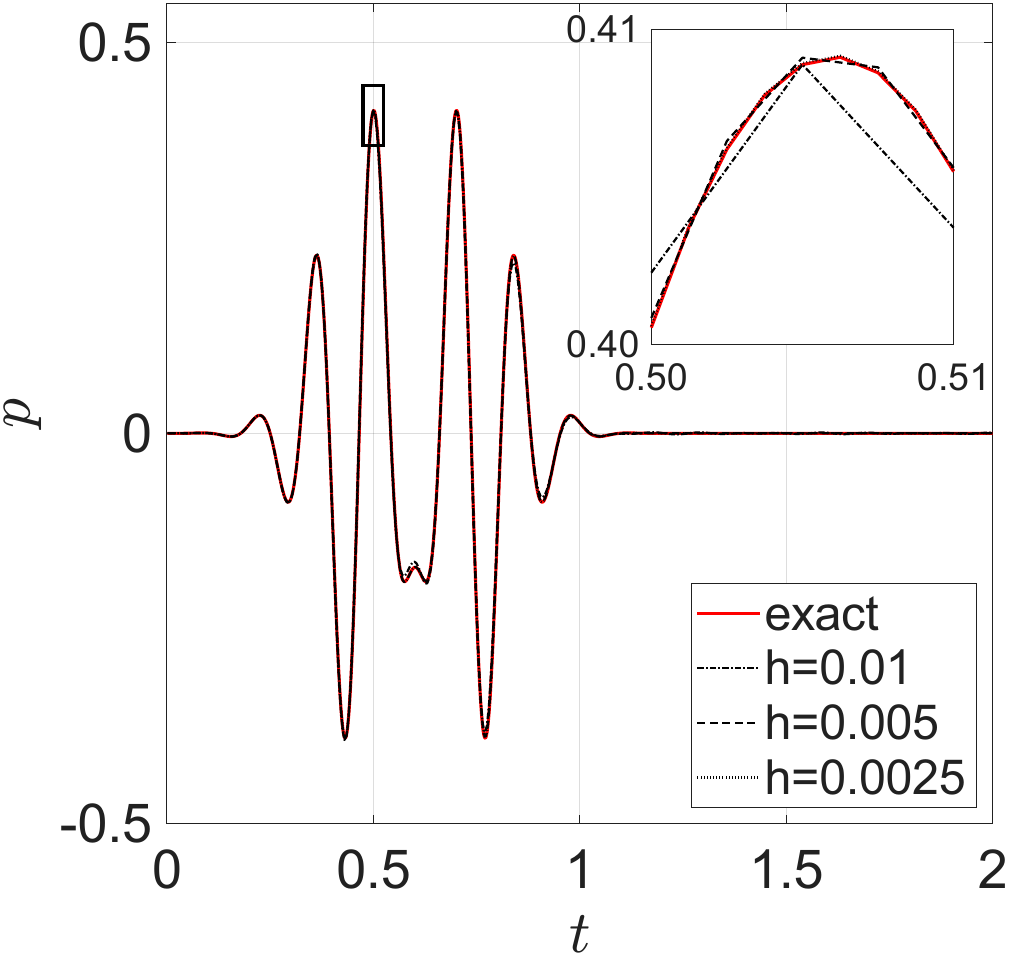}
        \caption{}
    \end{subfigure}
    \caption{Time histories of pressure fluctuation at $(0.5,0)$ for all the grids : (a) impedance wall and (b) rigid wall.}
    \label{fig:1d-micro}
\end{figure}
To verify whether the ADE-IBM preserves the fourth-order accuracy of the numerical scheme, a convergence study is performed with mesh sizes $h = 0.01,\, 0.005,\, 0.0025, \, 0.00167$ and a constant Courant–Friedrichs–Lewy (CFL) number of 0.5. Figure~\ref{fig:1d-micro} compares the numerical and semi-analytical solutions for rigid and impedance walls at a representative microphone located at $(0.5, 0)$, a point common to all grids.
\begin{figure}[ht!]
\centering
\includegraphics[width=0.4\linewidth]{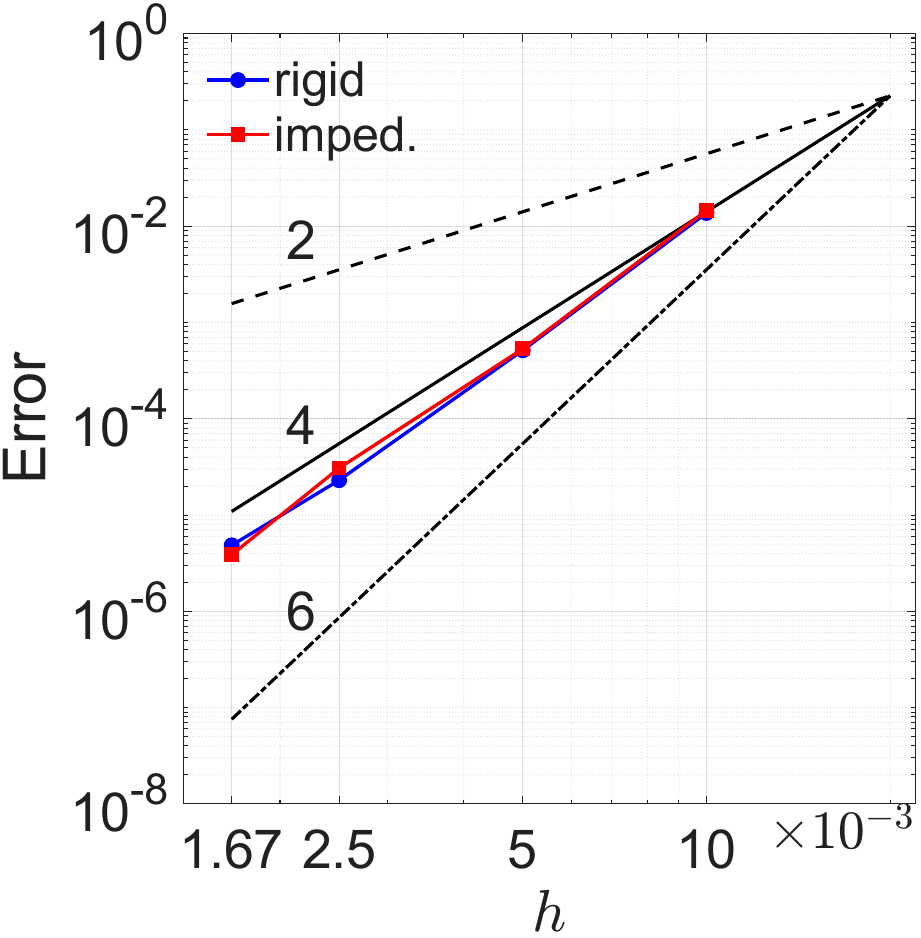}
\caption{Convergence of the error with mesh size.}
\label{fig:1d-conv}
\end{figure}
For each case, the computed solution shows good agreement with the reference solution and converges toward it as $h$ decreases. The error is quantified as the root-mean-square (RMS) of the difference between the numerical and reference solutions over the recorded time window. The error is plotted versus the mesh size in Fig.~\ref{fig:1d-conv}. The order of convergence, indicated by the slope of the curves, is 4, both for the rigid and impedance walls. This is the desired order, since the reconstruction chain for the GPs was build to reach a formal order of $4$.

\subsection{Cylinder reflecting an acoustic pulse}
\label{sec:pulse}

The second benchmark involves a curved surface: the reflection of an initial Gaussian pressure pulse by a cylinder immersed in a uniform Cartesian grid. This 2D configuration has previously been used to validate IBMs \cite{SeoMittal2011, SUN2012} for rigid walls. Here, the cylinder wall has a prescribed impedance. To the best of the authors' knowledge, no reference solution exists for this case; a semi-analytical solution is therefore derived in the appendix.

\begin{figure}[ht!]
    \centering
    \begin{subfigure}[t]{0.32\textwidth}
        \centering
        \includegraphics[width=\linewidth]{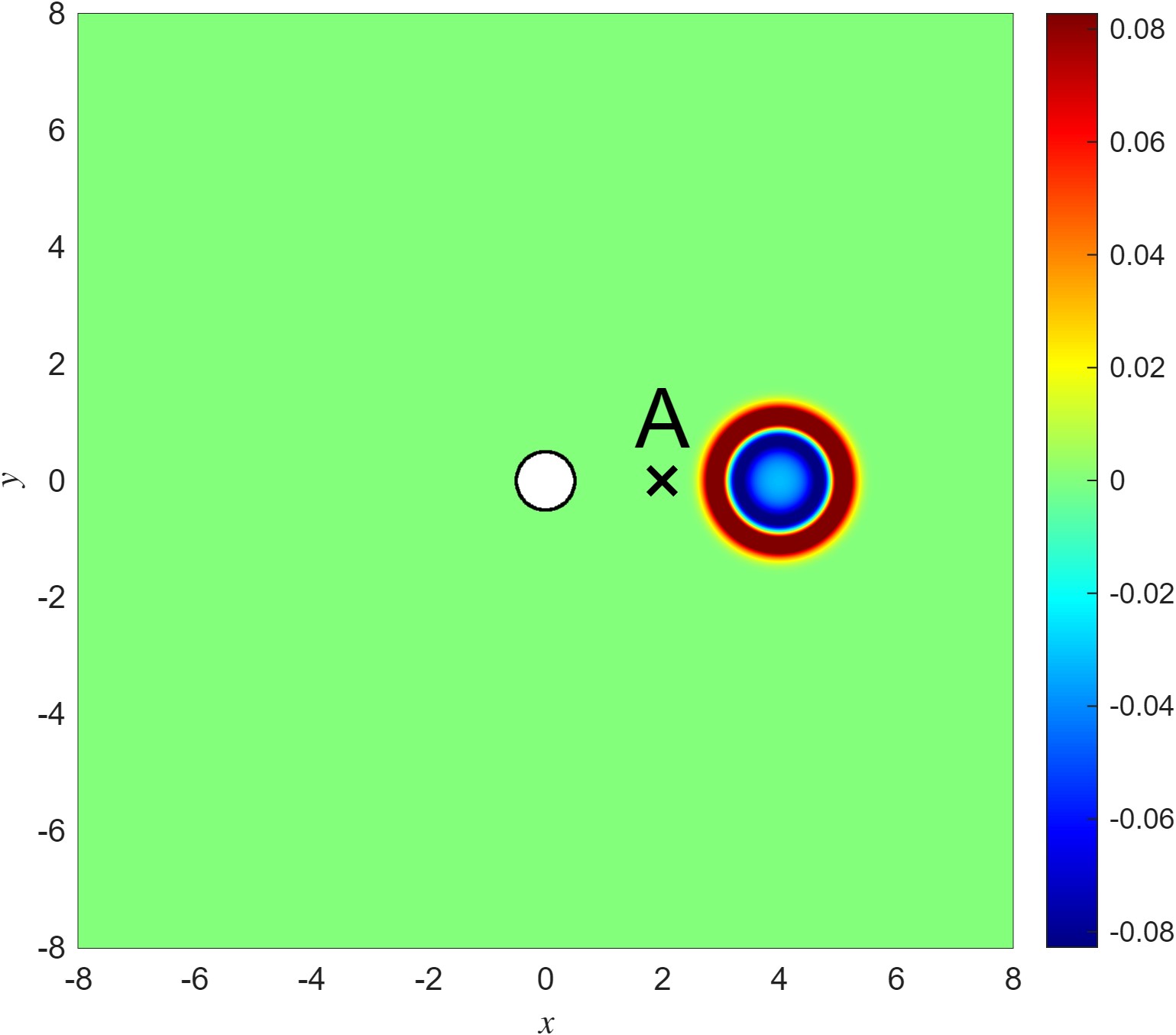}
        \caption{$t=1$}
    \end{subfigure}%
    ~ 
    \begin{subfigure}[t]{0.32\textwidth}
        \centering
        \includegraphics[width=\linewidth]{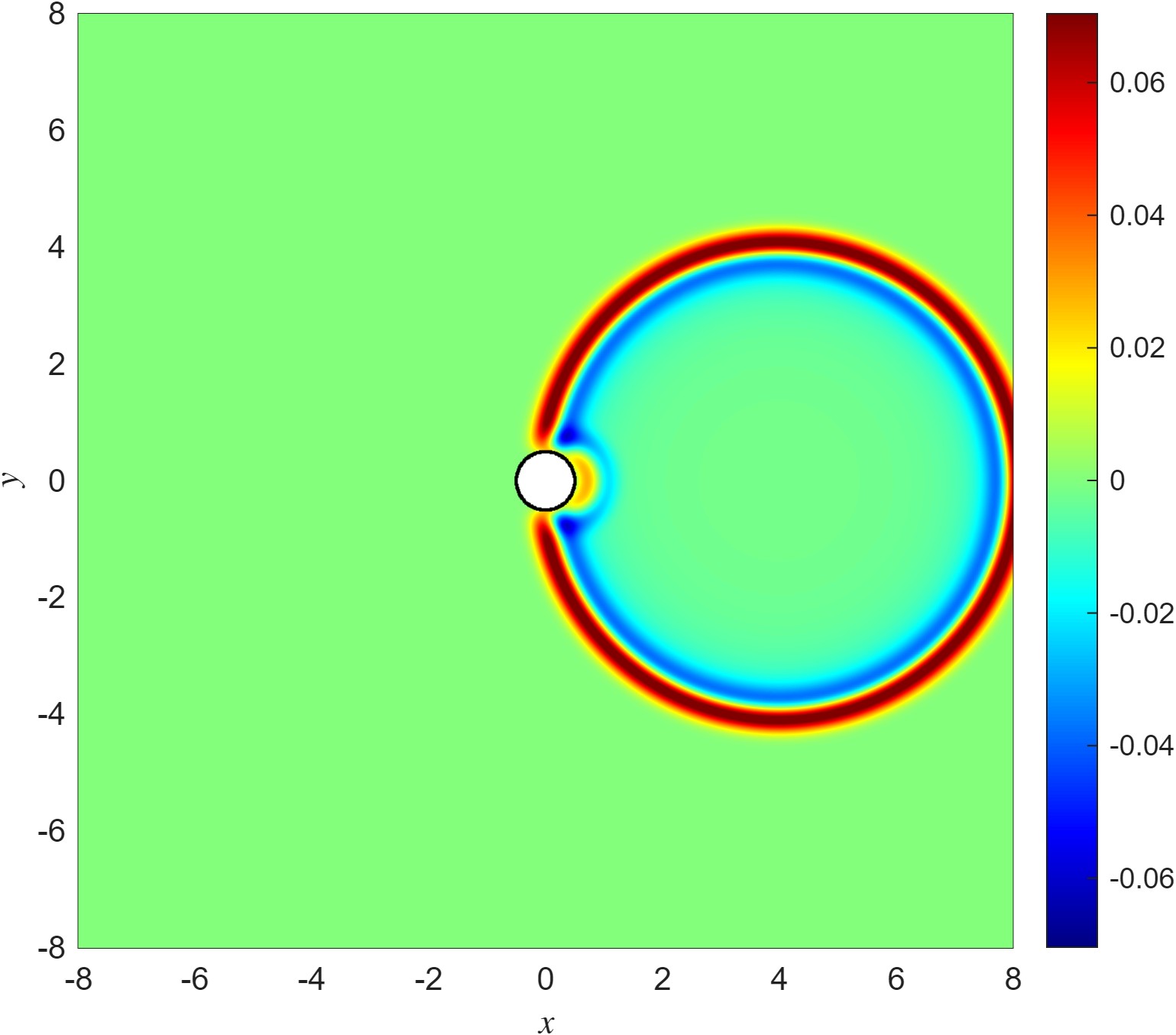}
        \caption{$t=4$}
    \end{subfigure}
    ~ 
    \begin{subfigure}[t]{0.32\textwidth}
        \centering
        \includegraphics[width=\linewidth]{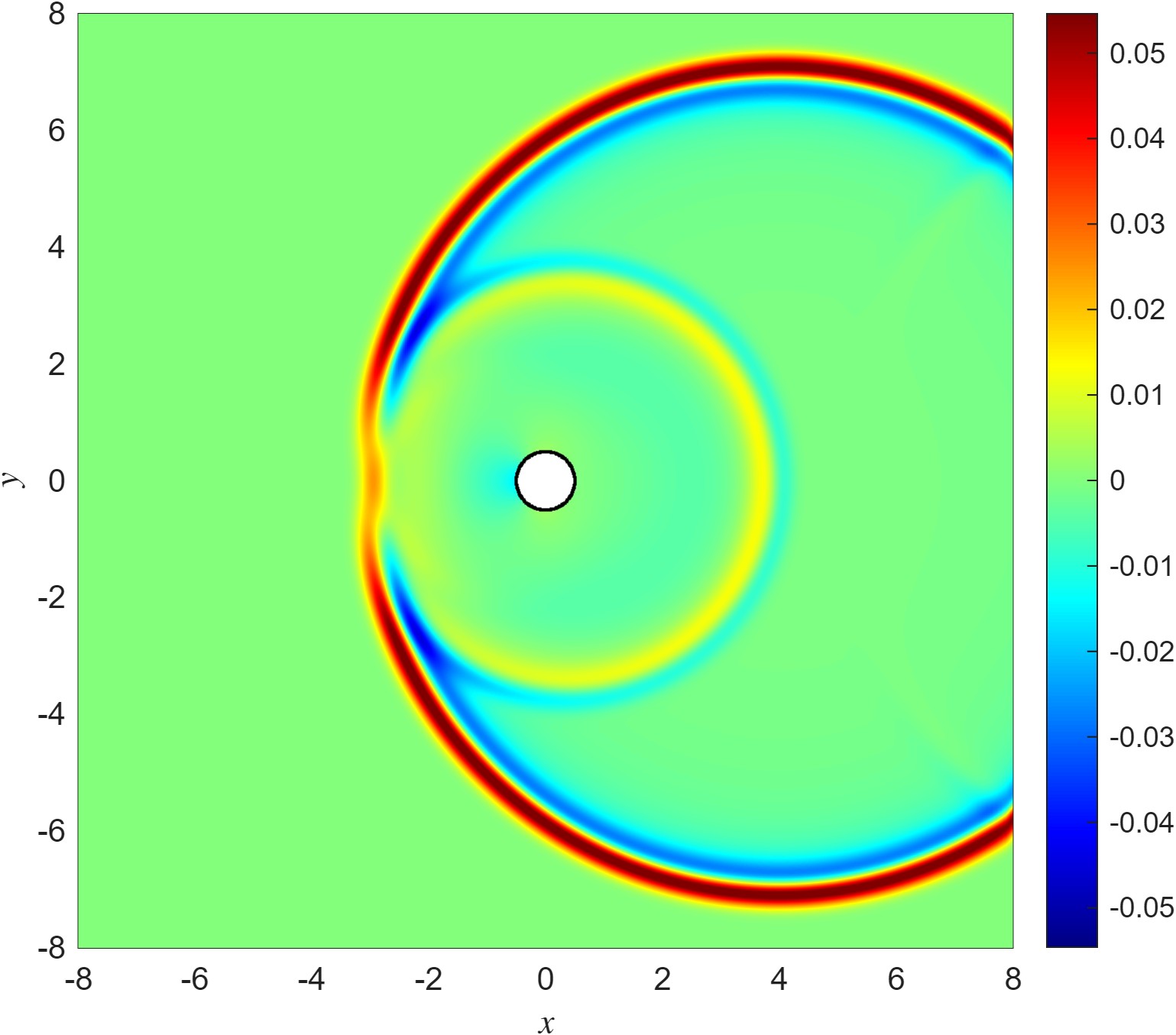}
        \caption{$t=7$}
    \end{subfigure}
    \caption{Pressure snapshots at different instants during the diffraction of a pulse by a cylinder with impedance.}
    \label{fig:pulse-field}
\end{figure}

\begin{figure}[ht!]
    \centering
    \begin{subfigure}[t]{0.5\textwidth}
        \centering
        \includegraphics[width=\linewidth]{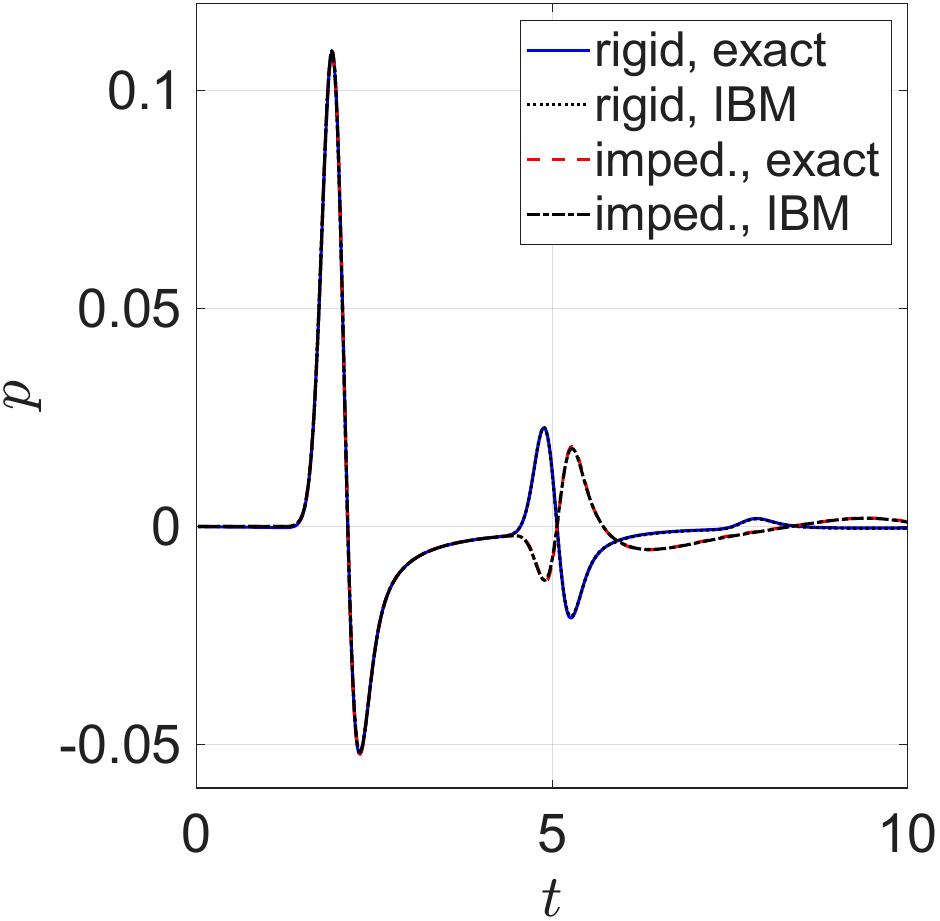}
        \caption{}
    \end{subfigure}%
    ~ 
    \begin{subfigure}[t]{0.5\textwidth}
        \centering
        \includegraphics[width=\linewidth]{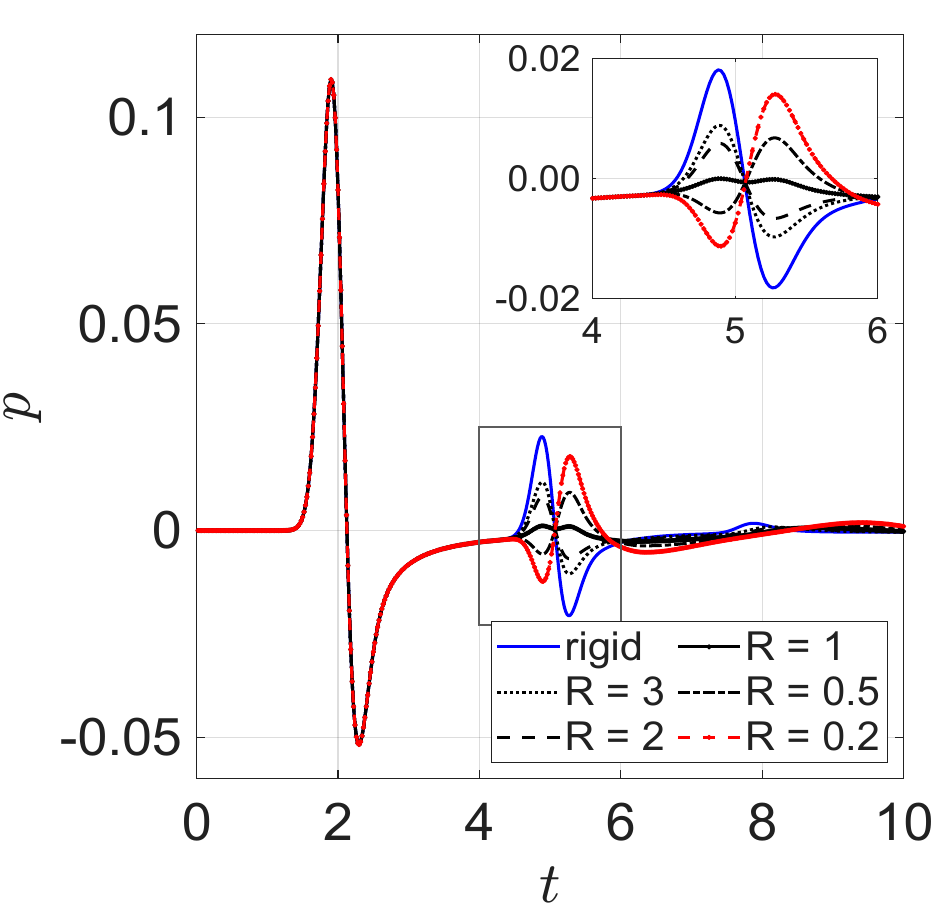}
        \caption{}
    \end{subfigure}
    \caption{Time history of pressure pressure at point A $(2,0)$: (a) rigid and impedance wall simulations compared with the semi-analytical solution; (b) rigid and impedance wall simulations for several values of $R_m\in\{3,2,1,0.5,0.2\}$.}
    \label{fig:pulse-resist}
\end{figure}

The computational domain is $[-6,6]\times[-6,6]$, 
discretized on a uniform Cartesian grid, with
$h=0.04$ and a CFL number of $0.5$, i.e.\ $\Delta t=0.02$.
The cylinder, of radius $r_c=0.5$, is centered at the origin, and, as before, the three layers of ghost points inside it are reconstructed by the ADE-IBM of Sec.~\ref{sec:immersedboundary}. NRBCs are imposed on all four sides
of the domain. The initial pulse at $t=0$, centered at $(x_0,y_0)=(4,0)$, satisfies
\begin{equation}
p_0(x,y) = a\, e^{-b\left[(x-x_0)^2+(y-y_0)^2\right]}, \qquad
u_0 = v_0 = 0,
\end{equation}
where the amplitude is $a=1$ and $b = \frac{\ln 2}{w^2}$ with $w=0.2$, the initial pulse width. The cylinder wall is either rigid or has an admittance given by Eq.~(\ref{eq:MSD_admittance}) with $M_m=0.025$, $K_m=1.6$, $R_m=0.2$. The corresponding resonance frequency is $\omega_{\rm res} = \sqrt{K_m/M_m}= 8$, which lies within the low frequency range excited by the pulse. A general overview of the pulse reflection from the impedance wall is shown in Fig.~\ref{fig:pulse-field}, which displays pressure snapshots at three instants, $t=1$, $t=4$, and $t=7$, corresponding respectively to before, during, and after the interaction of the pulse with the cylinder. At $t=7$, the reflected pulse is seen, which is entirely determined by the interaction with the wall; its accuracy depends on the order of the IBM.
%
Figure~\ref{fig:pulse-resist}a compares the simulated pressure at point A, located at $(x,y)=(2,0)$, with the semi-analytical solution (given by Eq.~(\ref{eq:pulse_impedance_cylinder_solution})), for both rigid and impedance walls. At this point, the reflected pulse arrrives approximately in the time interval $t \in [4 \; 6]$. The computed solution shows good agreement with the reference for both wall conditions.
%
Figure~\ref{fig:pulse-resist}b illustrates the physical effect of the wall resistance by plotting the pressure at point A for several values of $R_m$, ranging from the rigid-wall limit down to $R_m=0.2$, with $K_m$ and $M_m$ held constant. This decrease swaps the positive and negative parts of the pulse: the tail corresponds to an overpressure for large $R_m$ (rigid case) and to a negative pressure for small $R_m$. For $R_m = 1$, the normal impedance of the cylinder equals the characteristic impedance of air, resulting in a minimal amplitude of the reflected pulse, as intuitively expected.

\begin{figure}[ht!]
\centering
\includegraphics[width=0.4\linewidth]{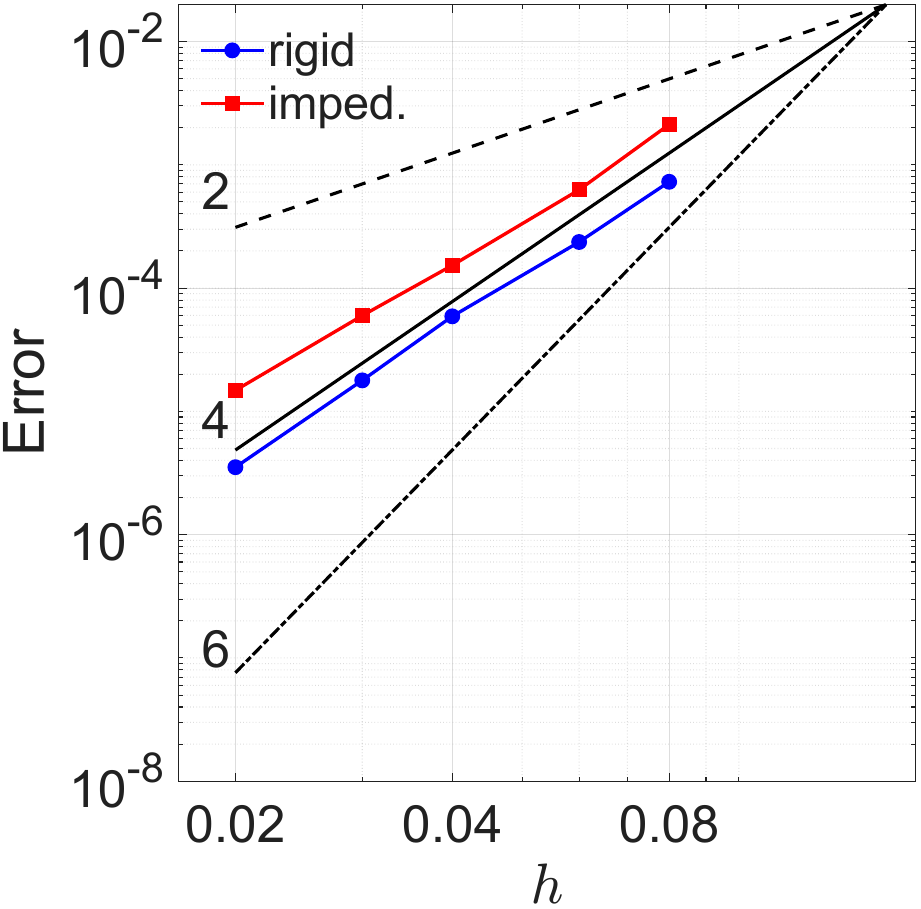}
\caption{Error on the reflected pulse versus mesh size.}
\label{fig:2d-conv_pulse}
\end{figure}

A convergence study is carried out by using five different values of the mesh size: $h=0.08 ,\, 0.06 ,\, 0.04 ,\, 0.03 ,\, 0.02$. The error is calculated against the reference provided the semi-analytical solution for both the rigid and impedance walls. The error is computed for the time window $t\in [4, 6]$ at the microphone $(2,0)$, which contains the information of the interaction of the pulse with the wall. For both the rigid and the impedance wall, the simulation converges toward the reference, as seen in Figure~\ref{fig:2d-conv_pulse}. The slopes show an order close to $4$ for the rigid wall, and a slightly decreased order of 3.6 for the impedance wall.
%
Overall, the high-order accuracy of the solver is preserved with the ADE-IBM.

\subsection{Cylinder scattering a harmonic monopole}
\label{sec:diffraction}
\begin{figure}[ht]
    \centering
    \includegraphics[width=0.5\linewidth]{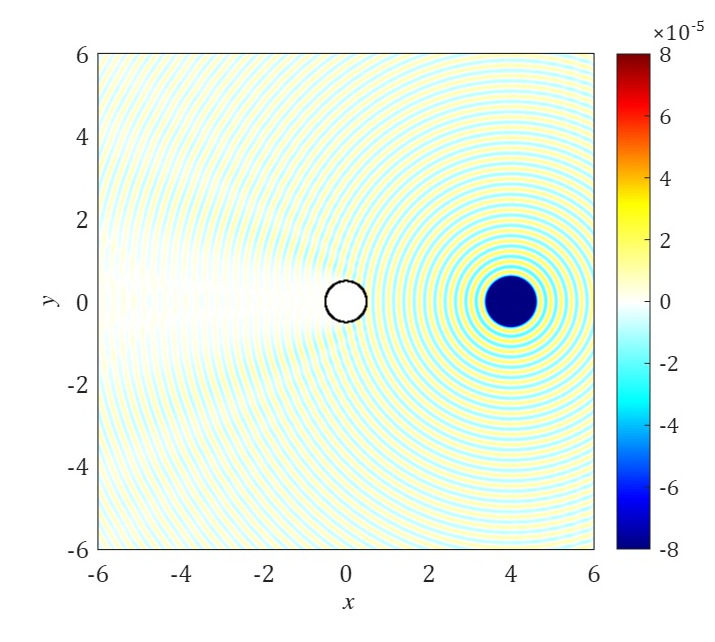}
    \caption{Cylinder with impedance wall scattering a harmonic monopole: instantaneous pressure field in harmonic steady state ($t=15$).}
    \label{fig:diff-field}
\end{figure}

\begin{figure}[ht!]
    \centering
    \begin{subfigure}[t]{0.5\textwidth}
        \centering
        \includegraphics[width=\linewidth]{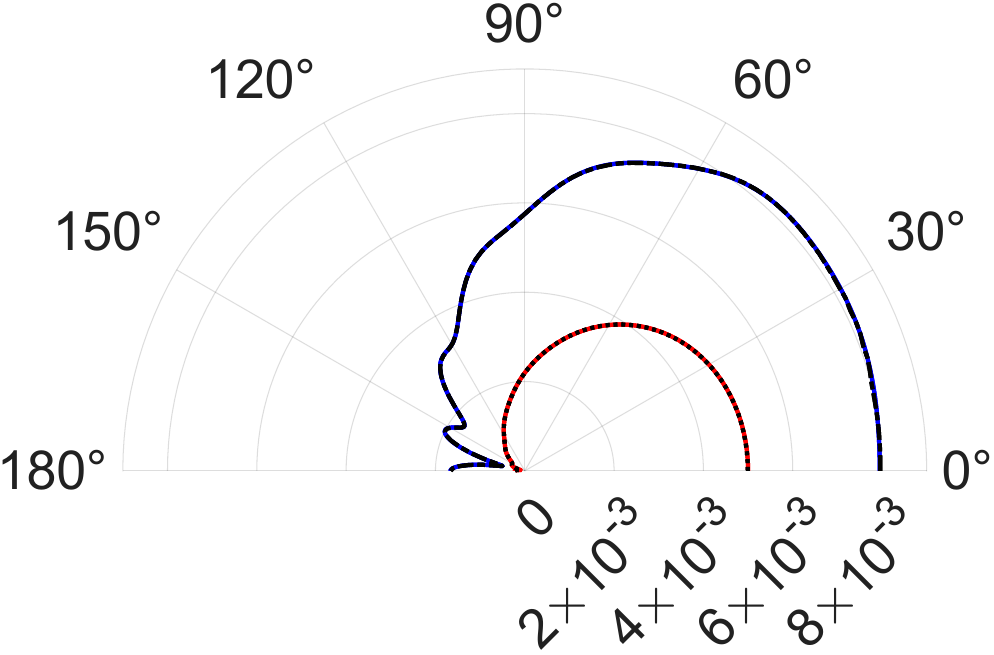}
        \caption{$r=0.55$}
    \end{subfigure}%
    ~ 
    \begin{subfigure}[t]{0.5\textwidth}
        \centering
        \includegraphics[width=\linewidth]{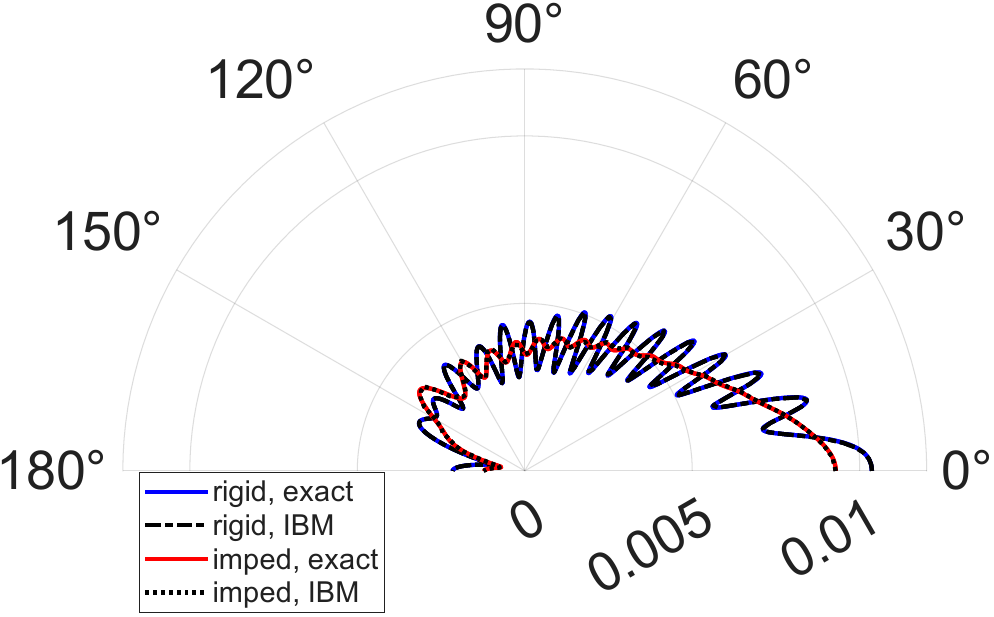}
        \caption{$r=5$}
    \end{subfigure}
    \caption{Cylinder scattering of a harmonic monopole: RMS scattered pressure compared with the exact solution, for both rigid and impedance walls, at radii (a) $r=0.55$ and (b) $r=5$.}
    \label{fig:diff-far}
\end{figure}

Having established the convergence order of the ADE-IBM, a final benchmark is now considered: the diffraction of a monopole sound field by a cylinder. This configuration is classical for testing IBMs with rigid walls \cite{SeoMittal2011} and is used here for an impedance wall. While the construction of the method focused on its order of accuracy, its stability was not demonstrated, as is sometimes possible\cite{BREHM2015, Bilbao2022JASA}. Hence, this test case also aims to verify the long-term stability of the method by simulating a large number of acoustic periods.\\
The case is close to the one of the previous section. The computational domain and the cylinder are unchanged, the mesh size is $h = 0.02$ and the time step is $\Delta t=0.01$, which corresponds to a CFL number of $0.5$. The admittance is given by Eq.~(\ref{eq:MSD_admittance}) with $M_m=0.025$, $K_m=15.79$, $R_m=1$. The resonance frequency is $\omega_{\rm res} = \sqrt{K_m/M_m} \sim 25$. A continuously forced monopole replaces the transient pulse of Sec.~\ref{sec:pulse}. In Eq.~(\ref{eq:LEE}), the source $s$ is: 
\begin{equation}
    s(x,y,t) = a\, e^{-b\left[(x-x_0)^2+(y-y_0)^2\right]} \sin(\omega t),
\end{equation}
where $a=1$, $b = \frac{\ln 2}{w^2}$, $w=0.2$, $(x_0, y_0) = (4,0)$, and the oscillation frequency, $\omega=8\pi$, matches $\omega_{\rm res}$.
The simulation is run for 60 periods, until $t_\mathrm{end}=15$, and reaches a harmonic steady state after $t=10$. Figure~\ref{fig:diff-field} shows an instantaneous scattered pressure field in steady state.
%
The RMS pressure calculated on the window $t\in[13,15]$, representing eight periods of forcing in steady state, is calculated along two circles of radii $r=0.55$ and $r=5$. This is shown in Fig.~\ref{fig:diff-far} for both rigid and impedance walls. An exact solution is also available and is included in the figure. The computed solution is again in good agreement with the exact solution in both the near and far fields. The stability of the method over a long period of time is thus verified. In addition, even though a filter would generally be applied with high-order methods\cite{SeoMittal2011} this simulation could be performed without a selective numerical filter at the walls. The least-squares fit probably participates in regularizing the solution.

%
%

\section{Conclusion}

An immersed boundary method coupled with an auxiliary-differential-equation
formulation of a locally-reacting impedance wall, referred to as ADE-IBM, has been developed for the linearized Euler equations without mean flow, solved with a high-order finite difference method. The IBM allows to account for walls non-aligned with the cartesian grid. It relies on a sharp interface method using ghost-points, which has the potential to preserve high-order precision in space. The reconstruction of the ghost points is based on a fourth-order normal stencil approach originally used for characteristic variables and rigid walls, but used here with primitive variables and rigid or impedance walls. The ADE method, whose objective is to determine the normal velocity to impose at the boundary points, has the potential to fit arbitrary impedance models without compromising the high-order temporal accuracy.

The method has been tested on the diffraction of an initial pressure pulse by both a flat wall and a cylinder, either rigid or having an impedance. The semi-analytical solution for the flat case is available in the literature, and that for a cylinder is developed in appendix. The order of convergence has been verified: it is generally 4, except for the cylinder with impedance, for which the order falls to 3.6. Overall, the high-order nature of the method is considered as preserved by the proposed method.
As high-order methods can be prone to instability close to boundaries, the long-term stability has been verified with the diffraction of a monopolar sound field by a cylinder over a large number of excitation periods. 

\section*{Appendices}
\appendix
\section{Pulse scattered by a cylinder of given impedance}
\label{sec:analytical_reflexion}
In this appendix, we present the semi-analytical solution for the scattering of an initial pressure pulse by a cylinder with a given acoustic impedance. The propagation of an initial Gaussian pressure pulse was given by Tam and Webb \citep{TAM1993}, and the solution for its scattering by on a rigid cylinder has been presented by Kurbastkii\citep{KURBATSKII1997}. An extension  to a cylinder with a given impedance is now presented. The method does not follow exactly Kurbastkii\citep{KURBATSKII1997} since the latter uses implicitely a Fourier transform, which may not be appropriate for an initial value problem. A unilateral Laplace transform (LT) is prefered, defined in Eq.~(\ref{eq:LT}). The same benchmark problem as Kurbatskii\citep{KURBATSKII1997} is considered, except here the cylinder has a locally reacting surface, specified by its admittance $Y(s)$. The cylinder of radius $a=0.5$ is placed at the origin of the coordinate system, $(x,y)$, or also $(r, \theta)$. The initial pressure pulse has a Gaussian distribution $p_0(x,y) = e^{-b((x-x_0)^2 + (y-y_0)^2)}$ centered at point $(x_0,y_0)=(4,0)$, and a nul velocity. This is also $p_0(r,\theta) = e^{-b((r \cos \theta-x_0)^2 + (r \sin \theta-y_0)^2)}$. The acoustic potential $\phi(r,\theta,t)$ satisfies $\Delta \phi - \frac{1}{c_a^2} \frac{\partial^2 \phi}{\partial t^2} = 0$. The acoustic pressure and velocity are respectively given by $p=-\rho_a \frac{\partial \phi}{\partial t}$ and ${\bf v} = \nabla \phi$. The initial conditions are $\nabla \phi)_{t=0} = {\bf v})_{t=0} = {\bf 0}$ and $\left.\frac{\partial \phi}{\partial t}\right)_{t=0} = -\frac{p_0}{\rho_a}$. The ambient speed of sound and density are $c_a=1$ and $\rho_a=1$. On the cylinder surface, at $r=a$, the impedance relation needs to be imposed, and the acoustic field is required to vanish in the far field.\\
By taking the LT of the wave equation for the potential, using $\hat{p} = -\rho_a s \hat{\phi}$, and injecting a Fourier series in azimut, $\hat{p}(r,\theta, s) = \sum_{n=-\infty}^\infty \hat{p}_{n}(r, s) e^{i n \theta}$, the following equation is obtained for the amplitude of the $n$-th azimuthal mode:
\begin{equation}
\label{eq:wave_eq_pn}
\frac{\partial^2 \hat{p}_n}{\partial r^2} + \frac{1}{r}\frac{\partial \hat{p}_n}{\partial r}   - \left( \frac{n^2}{r^2} + \frac{s^2}{c_a^2}\right) \hat{p}_n = \frac{-s}{c_a^2} p_{0,n}
\end{equation}
This is an inhomogenous modified Bessel equation. The two linearly independent solutions of the homongenous part are $I_n(sr/c_a)$ and $K_n(sr/c_a)$, where $I_n$ and $K_n$ are the modified Bessel functions of first and second kind. The inhomogeneous term, $p_{0,n}$, is the $n$-th azimuthal mode of the initial pressure pulse written as a Fourier series: $p_0(r, \theta) = \sum_{n=-\infty}^\infty p_{0,n}(r) e^{i n \theta}$. One can calculate $p_{0,n}(r) = \frac{1}{2\pi} \int_0^{2\pi} p_0(r,\theta) e^{-i n \theta} d\theta = e^{-b(r^2+ x_0^2)} I_n(2b r x_0)$. Equation~(\ref{eq:wave_eq_pn}) is solved as a scattering problem: the pressure is split into incident and scattered fields, $p(r,\theta, t) = p_{inc}(r, \theta,t) + p_{scat}(r, \theta, t)$, with a corrresponding Fourier Laplace decomposition $\hat{p}_n(r,s) = \hat{p}_{inc,n}(r, s) + \hat{p}_{scat,n}(r, s)$. The incident field should satisfy the problem without the cylinder, and the scattered field should be such that the total field satisfies the full problem, including the impedance boundary condition at the cylinder.\\
The incident field is first considered. It should satisfy  Eq.~(\ref{eq:wave_eq_pn}), without the cylinder, and thus be bounded at $r=0$. This can be solved by finding the Green's function $G_n(r,r')$ solution to:
\begin{equation}
\label{eq:wave_eq_pinc_n}
\frac{d^2 G_{n}}{d r^2} + \frac{1}{r}\frac{d G_{n}}{d r} - \left( \frac{n^2}{r^2} + \frac{s^2}{c_a^2}\right) G_{n} = \frac{-\delta(r-r')}{r}
\end{equation}
$G_n$ is a linear combination of $I_n$ and $K_n$ on each of the intervals $r<r'$ and $r>r'$ with a jump condition of its derivative at $r=r'$. Using the fact that $I_n$ is regular at $r=0$ and unbounded as $r\rightarrow\infty$, and that $K_n$ is bounded as $r\rightarrow\infty$, $G_n$ is given by:
\begin{equation}
G_n(r,r') = I_n\left(\frac{s}{c_a}r_<\right) K_n\left(\frac{s}{c_a}r_>\right)
\end{equation}
where $r_< = \mbox{min}(r,r')$ and $r_> = \mbox{max}(r,r')$.
This leads to the solution of the incident field:
\begin{equation}
    \hat{p}_{inc,n} (r,s) = \frac{s}{c_a^2} \int_0^\infty G_n(r,r') p_{0,n}(r') r'dr'
\end{equation}
This can be explicitely written as:
\begin{equation}
\label{eq:pinc_n}
\begin{split}
    \hat{p}_{inc,n} (r,s) = \frac{s}{c_a^2} K_n\left( \frac{sr}{c_a}\right) \int_0^r I_n\left( \frac{sr'}{c_a}\right) p_{0,n}(r') r' dr'\\ +  \frac{s}{c_a^2} I_n\left( \frac{sr}{c_a}\right) \int_r^\infty K_n\left( \frac{sr'}{c_a}\right) p_{0,n}(r') r' dr' 
    \end{split}
\end{equation}
Below, the partial derivative with respect to $r$ will be needed, it reads:
\begin{equation}
    \label{eq:dpincdr_n}
    \begin{split}
    \frac{\partial \hat{p}_{inc,n}}{\partial r} (r,s)= \frac{s^2}{c_a^3} K_n'\left( \frac{sr}{c_a}\right) \int_0^r I_n\left( \frac{sr'}{c_a}\right) p_{0,n}(r') r' dr'\\ +  \frac{s^2}{c_a^3} I_n'\left( \frac{sr}{c_a}\right) \int_r^\infty K_n\left( \frac{sr'}{c_a}\right) p_{0,n}(r') r' dr' 
    \end{split}
\end{equation}
where the prime on $I_n$ and $K_n$ indicates a derivative with respect to their argument.\\
Turning to the scattered field, it should satisfy Eq.~(\ref{eq:wave_eq_pn}) without the right hand side. In addition, for the scattered field to be bounded as $r \rightarrow\infty$, one should have $\hat{p}_{scat,n}(r,s) = A_n(s) K_n\left( \frac{sr}{c_a}\right)$. The mode amplitudes, $A_n(s)$, should be such that the total $\hat{p}$ field satisfies the impedance boundary condition at the cylinder surface, at $r=a$, that is $\hat{v}_r(s) = - Y(s) \hat{p}(s)$, where $\hat{v}_r = \frac{\partial\hat{\phi}}{\partial r}$. This provides :
\begin{equation}
\label{eq:mode_BC}
\left. \frac{\partial \hat{p}_{n}}{\partial r} \right)_{r=a} (a,s) = \rho_a s Y(s) \hat{p}_n(a,s)
\end{equation}
Injecting the total field $\hat{p}_n(r,s) = \hat{p}_{inc,n}(r,s) + \hat{p}_{scat,n}(r,s) = \hat{p}_{inc,n}(r,s) + A_n(s) K_n\left( \frac{sr}{c_a}\right)$ in the latter gives finally:
\begin{equation}
\label{eq:An}
A_n(s) = \frac{\rho_a s Y(s) \hat{p}_{inc,n}(a,s) - \frac{\partial  \hat{p}_{inc,n}}{\partial r}(a,s)}{\frac{s}{c_a}K_n'\left( \frac{sa}{c_a}\right) - \rho_a s Y(s) K_n\left( \frac{sa}{c_a}\right)}
\end{equation}
The full solution is obtained as follows. First, $\hat{p}_{inc,n}(r,s)$ is computed by numerical integration of Eq.~(\ref{eq:pinc_n}). Its value, and that of its derivative, Eq.~(\ref{eq:dpincdr_n}), are evaluated at $r=a$ and injected into Eq.~(\ref{eq:An}), which gives $A_n(s)$, and hence  $\hat{p}_{scat,n}(r,s)$. The sum of $\hat{p}_{inc,n}$ and $\hat{p}_{scat,n}$ gives the total $\hat{p}_{n}(r,s)$. The summation of the Fourier series over $n$ gives $\hat{p}(r,\theta,s)$. Finally, the pressure $p(r, \theta, t)$ is obtained by computing the inverse Laplace transform:
\begin{equation}
    \label{eq:pulse_impedance_cylinder_solution}
    p(r,\theta, t) = \frac{1}{2i\pi}\int_\Gamma \hat{p}(r, \theta, s) e^{s t} ds
\end{equation}
where $\Gamma =\{s; s = \alpha+i\omega, -\infty < \omega < \infty\}$ is a  Bromwich contour, and $\alpha$ is real positive. If only the incident field is considered (by enforcing $A_n(s)=0$), the present solution recovers the solution in \citet{TAM1993}, obtained in a different way.


\bibliographystyle{unsrtnat}
\bibliography{biblio}

@article{Bilbao2022JASA,
author={Bilbao, S.},
year={2022},
title={Immersed boundary methods in wave-based virtual
 acoustics},
volume={151},
journal={J. Acoust. Soc. Am.},
doi = {10.1121/10.0009768},
pages = {1627--1638}
}

@article{Bilbao2023JASA,
author={Bilbao, S.},
year={2023},
title={Modeling impedance boundary conditions and acoustic barriers using the immersed boundary method: The one-dimensional case},
volume={153},
number={4},
journal={J. Acoust. Soc. Am.},
doi={10.1121/10.0017763},
pages={2023--2036}
}

@article{BIN2009,
author = {Bin, Jonghoon and Yousuff Hussaini, M. and Lee, Soogab},
title = {Broadband impedance boundary conditions for the simulation of sound propagation in the time domain},
journal = {J. Acoust. Soc. Am.},
volume = {125},
number = {2},
pages = {664-675},
year = {2009},
month = {02},
doi = {10.1121/1.2999339}
}

@article{BLUMRICH2002,
author = {Blumrich, Reinhard and Heimann, Dietrich},
title = {A linearized Eulerian sound propagation model for studies of complex meteorological effects},
journal = {J. Acoust. Soc. Am.},
volume = {112},
number = {2},
pages = {446-455},
year = {2002},
month = {07},
doi = {10.1121/1.1485971}
}

@article{Bocquet2023JTCA,
author={Bocquet, C. and Desjouy, C. and Gabard, G.},
year={2023},
title={A high-order immersed moving boundary method using ghost points and characteristics for acoustics},
volume={32},
number={2}, 
journal={J. Theor. Comput. Acoust.},
doi={10.1142/S2591728523300015},
pages={2330001}
}

@article{BogeyBailly2004,
author={Bogey, C. and Bailly, C.},
year={2004},
title={A family of low dispersive and low dissipative explicit schemes for flow and noise computations},
volume={194},
number = {1},
journal={J. Comput. Phys.},
doi = {10.1016/j.jcp.2003.09.003},
pages = {194--214}
}

@article{BOTTELDOOREN1994,
title={Acoustical finite‐difference time‐domain simulation in a quasi‐Cartesian grid},
author={D. Botteldooren},
journal={J. Acoust. Soc. Am.},
year={1994},
volume={95},
pages={2313-2319},
doi = {10.1121/1.409866}
}

@article{BREHM2015,
title = {A locally stabilized immersed boundary method for the compressible \uppercase{N}avier–\uppercase{S}tokes equations},
journal = {J. Comput. Phys.},
author = {C. Brehm and C. Hader and H.F. Fasel},
volume = {295},
pages = {475-504},
year = {2015},
doi = {10.1016/j.jcp.2015.04.023}
}

@inproceedings{Diaz2022CFA,
author={M.~A. Diaz and V. Fortun{\'e} and D. Marx and C. Prax},
title={A high-order immersed boundary technique for computational aeroacoustics},
booktitle={16{\`e}me Congr{\`e}s Fran{\c c}ais d'Acoustique (CFA 2022), Marseille, France}, 
year={2022},
url = {hal.science/hal-03848118v1}
}

@article{DRAGNA2015,
author={D. Dragna and P. Pineau and Ph. Blanc-Benon},
year={2015},
title={A generalized recursive convolution method for time-domain propagation in porous media},
volume={138},
journal={J. Acoust. Soc. Am.},
doi = {10.1121/1.4927553},
pages = {1030--1042}
}

@article{FUNG2000,
title={Impedance and Its Time-Domain Extensions},
author={Ky-Y. Fung and Hong-Bin Ju and B. Phanindra Tallapragada},
journal={AIAA J.},
year={2000},
volume={38},
pages={30-38},
doi={10.2514/2.950}
}

@article{FUKUSHIMA2016,
author = {Fukushima, Y. and Misaka, T. and Obayashi, S. and Sasaki, D. and Nakahashi, K.},
title = {Wavenumber Optimized Immersed Boundary Method for Aeroacoustic Analysis Based on Cartesian Mesh},
journal = {AIAA J.},
volume = {54},
number = {10},
pages = {2988-3001},
year = {2016},
doi = {10.2514/1.J053958}
}

@ARTICLE{GUSTAVSEN1999,
author={Gustavsen, B. and Semlyen, A.},
journal={IEEE Transactions on Power Delivery}, 
title={Rational approximation of frequency domain responses by vector fitting},
year={1999},
volume={14},
number={3},
pages={1052-1061},
doi={10.1109/61.772353}
}

@article{HEUTSCHI2005,
author = {Heutschi, K. and Horvath, Matthias and Hofmann, Jan},
year = {2005},
month = {01},
pages = {35-40},
title = {Simulation of Ground Impedance in Finite Difference Time Domain Calculations of Outdoor Sound Propagation},
volume = {91},
journal = {Acta Acustica united with Acustica}
}

@inproceedings{KURBATSKII1997,
author={K. A. Kurbastkii},
title={Analytical solutions for the category 1, benchmark problems 1 and 2},
booktitle={Second Computational Aeroacoustics (\uppercase{CAA}) workshop on benchmark problems},
year={1997},
editor={C. K. W. Tam and J. C. Hardin},
pages={9-14},
note={\uppercase{NASA} Conference Publication 3352}
}

@article{MARX2015,
author = {D. Marx},
title = {Numerical Computation of a Lined Duct Instability Using the Linearized \uppercase{E}uler Equations},
journal = {AIAA J.},
volume = {53},
number = {8},
pages = {2379-2388},
year = {2015},
doi = {10.2514/1.J053746}
}

@techreport{MITCHELL1995,
author      = {Mitchell, B. E. and Lele, S. K. and Moin, P.},
title       = {Direct Computation of the Sound Generated by Subsonic and Supersonic Axisymmetric Jets},
institution = {Center for Turbulence Research, Stanford University},
year        = {1995},
number      = {report TF-66},
month       = {November},
address     = {Stanford, CA}
}

@article{MITTAL2003,
title = {A versatile sharp interface immersed boundary method for incompressible flows with complex boundaries},
journal = {J. Comput. Phys.},
volume = {227},
number = {10},
pages = {4825-4852},
year = {2008},
doi = {10.1016/j.jcp.2008.01.028},
author = {R. Mittal and H. Dong and M. Bozkurttas and F.M. Najjar and A. Vargas and A. {von Loebbecke}}
}

@article{MITTAL_IACCARINO2005,
author = "Mittal, Rajat and Iaccarino, Gianluca",
title = "IMMERSED BOUNDARY METHODS", 
journal= "Annual Rev . Fluid Mech.",
year = "2005",
volume = "37",
number = "Volume 37, 2005",
pages = "239-261",
doi = "10.1146/annurev.fluid.37.061903.175743"
}

@book{OPPENHEIM_BOOK,
Author={A. V. Oppenheim and A. S. Willsky and S. H. Nawab},
Publisher = {Prentice Hall},
Title ={Signals and systems},
Edition={2nd},
Year={1997}
}

@phdthesis{Ouattara2025These,
author={Ouattara, A.},
title={D{\'e}veloppement d'une m{\'e}thode de fronti{\`e}res immerg{\'e}es en a{\'e}rocoustique num{\'e}rique. \uppercase{A}pplication {\`a} l'{\'e}coulement sur un r{\'e}seau de cavit{\'e}s
  acoustiques quarts d'onde},
type={Ph.{D}. Dissertation},
school={Universit{\'e}
  de Poitiers},
year={2025},
note={In French}
}

@misc{ouattara2026IBM,
title={High-order immersed boundary method for subsonic aeroacoustics and compressible \uppercase{N}avier-\uppercase{S}tokes equations}, 
year={2026},
eprint={2609.24729},
archivePrefix={arXiv},
primaryClass={physics.flu-dyn},
doi = {10.48550/arXiv.2609.24729},
author={A. Ouattara and D. Marx and C. Prax and V. Fortun{\'e} and M. Diaz and P. Jordan},
url={arxiv.org/abs/2609.24729}
}

@article{PESKIN1972,
author = {C. S. Peskin},
title = {Flow patterns around heart valves: A numerical method},
journal = {J. Comput. Phys.},
volume = {10},
number = {2},
pages = {252-271},
year = {1972},
doi = {10.1016/0021-9991(72)90065-4}
}

@article{reymen2008,
author = {Reymen, Y. and Baelmans, M. and Desmet, W.},
title = {Efficient Implementation of \uppercase{T}am and \uppercase{A}uriault's Time-Domain Impedance Boundary Condition},
journal = {AIAA J.},
volume = {46},
number = {9},
pages = {2368-2376},
year = {2008},
doi = {10.2514/1.35876}
}

@article{SALOMONS2002,
author = {Salomons, Erik and Blumrich, R. and Heimann, D.},
year = {2002},
month = {07},
pages = {483-492},
title = {Eulerian Time-Domain Model for Sound Propagation over a Finite-Impedance Ground Surface. \uppercase{C}omparison with Frequency-Domain Models},
volume = {88},
journal = {Acustica - Acta Acustica}
}

@article{SEBASTIAN2019,
title = {Numerical simulation of a turbulent channel flow with an acoustic liner},
journal = {J. Sound Vib.},
volume = {456},
pages = {306-330},
year = {2019},
doi = {https://doi.org/10.1016/j.jsv.2019.05.020},
author = {Robin Sebastian and David Marx and Véronique Fortun{\'e}}
}

@article{SeoMittal2011,
author = {Jung Hee Seo and Rajat Mittal},
title = {A high-order immersed boundary method for acoustic wave scattering and low-\uppercase{M}ach number flow-induced sound in complex geometries},
journal = {J. Comput. Phys.},
volume = {230},
number = {4},
pages = {1000-1019},
year = {2011},
issn = {0021-9991},
doi = {10.1016/j.jcp.2010.10.017}
}

@article{SUN2012,
author = {Sun, Xiaofeng and Jiang, Yongsong and Liang, An and Jing, Xiaodong},
title = {An immersed boundary computational model for acoustic scattering problems with complex geometries},
journal = {J. Acoust. Soc. Am.},
volume = {132},
number = {5},
pages = {3190-3199},
year = {2012},
doi = {10.1121/1.4757747}
}

@article{TAM1993,
title = {Dispersion-Relation-Preserving Finite Difference Schemes for Computational Acoustics},
journal = {J. Comp. Phys.},
volume = {107},
number = {2},
pages = {262-281},
year = {1993},
author = {Christopher K.W. Tam and Jay C. Webb},
doi = {10.1006/jcph.1993.1142}
}

@article{TAM1994,
author = {Tam, C. K. W. and Dong, Z.},
title = {Wall boundary conditions for high-order finite-difference schemes in computational aeroacoustics},
journal = {Theoret. Comput. Fluid Dynamics},
volume = {6},
pages = {303–322},
year = {1994},
doi = {10.1007/BF00311843}
}

@article{TAM1996,
author = {C. Tam and Z. Dong},
year = {1996},
month = {11},
pages = {175-201},
title = {Radiation and outflow boundary conditions for direct computation of acoustic and flow disturbances in a nonuniform mean flow},
volume = {04},
journal = {J. Comput. Acoustics},
doi = {10.1142/S0218396X96000040}
}

@article{TamAuriault1996,
author={Tam, C.~K.~W. and Auriault, L.},
year={1996},
title={Time-domain impedance boundary conditions for computational aeroacoustics},
volume={34},
number={5},
journal={AIAA J.},
doi = {10.2514/3.13167},
pages = {917--923}
}

@article{WILSON2007,
title = {Time-domain calculations of sound interactions with outdoor ground surfaces},
journal = {Appl. Acoust.},
volume = {68},
number = {2},
pages = {173-200},
year = {2007},
doi = {10.1016/j.apacoust.2005.10.004},
author = {D. Keith Wilson and Vladimir E. Ostashev and Sandra L. Collier and Neill P. Symons and David F. Aldridge and David H. Marlin},
}

@article{TROIAN2017,
author={R. Troian and D. Dragna and C. Bailly and M.-A. Galland},
year={2017},
title={Broadband liner impedance eduction for multimodal acoustic propagation in the presence of a mean flow},
volume = {392},
journal = {J. Sound and Vib.},
doi = {10.1016/j.jsv.2016.10.014},
pages = {200-216}
}

@article{ZHAO2021,
title = {A sharp interface immersed boundary method for flow-induced noise prediction using acoustic perturbation equations},
journal = {Computers and Fluids},
volume = {227},
pages = {105032},
year = {2021},
doi = {10.1016/j.compfluid.2021.105032},
author = {C. Zhao and Y. Yang and T. Zhang and H. Dong and G. Hou}
}


\end{document}